\documentclass[manuscript,screen]{acmart}

\AtBeginDocument{%
  \providecommand\BibTeX{{%
    \normalfont B\kern-0.5em{\scshape i\kern-0.25em b}\kern-0.8em\TeX}}}

\PassOptionsToPackage{printonlyused,smaller,nohyperlinks,nolist}{acronym}
\usepackage{acronym}
\usepackage{subcaption}
\graphicspath{ {figures/} }
\DeclareGraphicsExtensions{.jpg,.png,.pdf}
\usepackage{multirow}

\usepackage{tabularx}
\newcolumntype{L}[1]{>{\raggedright\arraybackslash}p{#1}}
\newcolumntype{C}[1]{>{\centering\arraybackslash}p{#1}}
\newcolumntype{R}[1]{>{\raggedleft\arraybackslash}p{#1}}

\newcommand{\ie}{i.\,e.~}

\newcommand{\eg}{e.\,g.~}

\newcommand{\cf}{c.\,f.~}

\begin{document}

\title{NxTF: An API and Compiler for Deep Spiking Neural Networks on Intel Loihi}


\author{Bodo Rueckauer}
\email{rbodo@ini.uzh.ch}
\orcid{0000-0003-1628-707X}
\affiliation{%
  \institution{
  Institute of Neuroinformatics, University of Zurich and ETH Zurich}
  \streetaddress{Winterthurerstrasse 190}
  \city{Zurich}
  \postcode{8057}
  \country{Switzerland}
  }

\author{Connor Bybee}
\email{bybee@berkeley.edu}
\affiliation{%
  \institution{
  Redwood Center for Theoretical Neuroscience, University of California, Berkeley}
  \streetaddress{575A Evans Hall, MC\# 3198}
  \city{Berkeley}
  \postcode{94720-3198}
  \country{USA}
  }

\author{Ralf Goettsche}
\email{ralf.goettsche@intel.com}
\affiliation{%
  \institution{Intel Corporation}
  \streetaddress{Am Campeon 10}
  \city{Neubiberg}
  \postcode{85579}
  \country{Germany}
  }

\author{Yashwardhan Singh}
\email{yashwardhan.singh@intel.com}
\author{Joyesh Mishra}
\email{joyesh.mishra@intel.com}
\author{Andreas Wild}
\email{andreas.wild@intel.com}
\affiliation{
  \institution{Intel Corporation}
  \streetaddress{2111 NE 25th Ave}
  \city{Hillsboro}
  \postcode{97124}
  \state{Oregon}
  \country{USA}
  }

\renewcommand{\shortauthors}{Rueckauer et al.}

\begin{abstract}
    \acfp{SNN} are a promising paradigm for efficient event-driven processing of spatio-temporally sparse data streams. \acp{SNN} have inspired the design and can take advantage of the emerging class of neuromorphic processors like Intel Loihi. These novel hardware architectures expose a variety of constraints that affect firmware, compiler and algorithm development alike. To enable rapid and flexible development of \ac{SNN} algorithms on Loihi, we developed NxTF: a programming interface derived from Keras and compiler optimized for mapping deep convolutional \acp{SNN} to the multi-core Intel Loihi architecture. We evaluate NxTF on \acp{DNN} trained directly on spikes as well as models converted from traditional \acp{DNN}, processing both sparse event-based and dense frame-based data sets. Further, we assess the effectiveness of the compiler to distribute models across a large number of cores and to compress models by exploiting Loihi's weight sharing features. Finally, we evaluate model accuracy, energy and time to solution compared to other architectures. The compiler achieves near optimal resource utilization of 80\% across 16 Loihi chips for a 28-layer, 4M parameter MobileNet model with input size $128\times128$. In addition, we report the lowest error rate of 8.52\% for the \acs{CIFAR}-10 dataset on neuromorphic hardware, using an off-the-shelf MobileNet.
\end{abstract}



\maketitle

\section{Introduction}

The widespread success of \ac{DL} has been accompanied by the development of user-friendly frameworks like Keras \citep{Chollet2015-lq} and PyTorch \citep{Paszke2019-ax} that ease the development and mapping of large scale machine learning models to state of the art \ac{DL} hardware. Still, training and running \acp{DNN} on conventional hardware is often costly in terms of time and energy consumption. This realization has driven research to develop more efficient algorithms (\eg \cite{Blalock2020-td}) and dedicated hardware (\eg \cite{Reuther2019-zj}). 

\acp{SNN} form a particular class of such algorithms that draw inspiration from the brain by transmitting information asynchronously in form of discrete spikes. Neuromorphic platforms like SpiNNaker \citep{Furber2013-ws}, TrueNorth \citep{Merolla2014-dt}, and Loihi \citep{Davies2018-xo} are optimized for this kind of event-based computation and have demonstrated the potential to run neural networks at much lower latency and energy consumption \citep{Stromatias2015-kl,Esser2016-ub,Davies2020} than traditional \acp{CPU} or \acp{GPU} based computer architectures. 

However, none of the current mainstream \ac{DL} software frameworks support these neuromorphic architectures for the purposes of \ac{DL} applications. Instead, the available neuromorphic software frameworks often expose the user to some of the complexities and constraints of these novel hardware architectures, therefore rendering them less accessible to end-users not familiar with neuromorphic technologies. There are several software frameworks for programming neuromorphic systems today. For instance, Nengo \citep{Bekolay2014-zm} is a cross-platform tool suite to train, simulate and map \acp{SNN} to different neuromorphic hardware platforms. PyNN \citep{Davidson2009-qh} is a simulator-independent programming language for neural networks that supports running the models on SpiNNaker and BrainScaleS \citep{Schmitt2017-cq}. In addition, each neuromorphic platform usually comes with its own set of programming abstractions, compiler, and firmware like Spynnaker \citep{Rhodes2018-wp}, SpiNNTools \cite{Rowley2019-qf}, TrueNorth Corelets \citep{Amir2013-fr} or Intel's NxSDK for the Loihi architecture.

\begin{figure}[htb]
    \centering
    \includegraphics[width=0.9\linewidth]{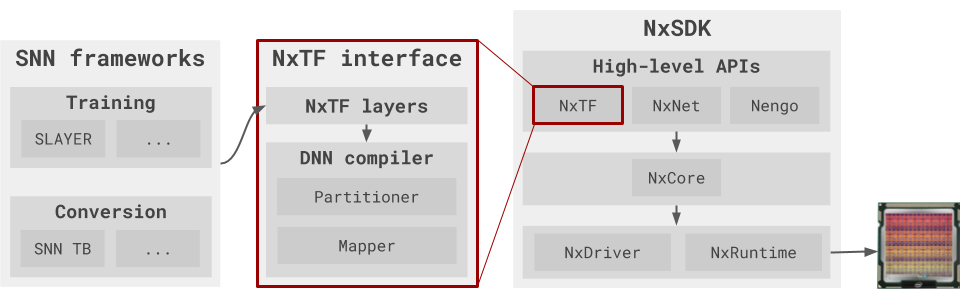}
    \caption{NxSDK software stack and workflow to configure a deep \ac{SNN} on Loihi. Starting from a trained or converted \ac{SNN}, the user defines a model in Python using the Keras-derived NxTF interface. The network is then partitioned and mapped onto Loihi by the \ac{DNN} compiler via the register-level NxCore \ac{API}. The NxDriver and NxRuntime components are responsible for interaction and execution of the \ac{SNN} on Loihi.}
    \label{fig:nxtf_workflow}
\end{figure}

Intel's NxSDK (see Fig.~\ref{fig:nxtf_workflow}) supports different high-level \acp{API}, each intended for different use cases, including the third-party \ac{API} Nengo and Intel's own \acp{API} NxNet as well as the new NxTF framework introduced here. All interfaces provide their own compiler to map a neural network definition to the lower-level NxCore \ac{API} that allows to configure a system of Loihi chips at the register level. NxNet provides full access to all computational features offered by Loihi. It allows to define neural networks at a structural level of individual cells or populations of neurons connected via axons and synapses that might be subject to synaptic learning rules. In contrast, NxTF trades generality for an application focus towards deep \acp{SNN}. This objective is achieved by inheriting from the Keras \textit{Model} and \textit{Layer} interface and providing a specialized \ac{DNN} compiler. To compress large \acsp{DNN} on Loihi, this compiler exploits the regular structure of typical \ac{DNN} topologies like \acp{CNN} in combination with Loihi's ability to share connectivity state. The NxTF source code is publicly available online\footnote{\url{https://github.com/intel-nrc-ecosystem/models/tree/master/nxsdk_modules_ncl/dnn}}.

We begin this paper by outlining the typical workflow of using the proposed NxTF framework to deploy a \ac{DNN} on Loihi in Sec.~\ref{sec:nxtf_overview}. After reviewing the necessary background about the Loihi architecture in Sec.~\ref{sec:loihi}, we explain the network compilation procedure in Sec.~\ref{sec:compilation}. To evaluate the effectiveness of the compiler, we measure the memory utilization when deploying architectures of different structure and size on Loihi (Sec.~\ref{sec:core_utilization}). Low memory utilization is essential on a non-Von-Neumann architecture like Loihi, where multiple smaller blocks of memory are distributed locally close to the computation. We also study how well NxTF is able to exploit Loihi's ability to share synapses and axons when compiling convolution layers in order to compress redundant connectivity information (Sec.~\ref{sec:sharing_evaluation}). In terms of applications, we present two common use cases of NxTF in Sec.~\ref{sec:nxtf_applications}: Deploying an \ac{SNN} on Loihi that a) has been trained directly on a spike-based dataset using \acs{SLAYER} \citep{Shrestha2018-io}, and b) that has been converted from an \ac{ANN} trained on frames. Finally, we benchmark the performance of these networks and compare their energy consumption and execution time against results on other neuromorphic hardware.

\section{Methods}

As our core contribution, we present here a user interface and compiler for \acp{DNN} on Loihi, called \emph{NxTF}. In the design of this compiler, we followed two objectives: (1) To provide an easy-to-use tool that adopts Keras' level of abstraction. (2) To optimize the allocation of resources on the neurocores and exploit Loihi's connection sharing features for efficient deployment of large-scale \acp{CNN}. With regard to the first objective, we will first provide a high-level overview of the tool chain in Sec.~\ref{sec:nxtf_overview}. We then introduce basic concepts of the Loihi hardware architecture to understand resource constraints and opportunities for resource sharing taken into account by the compiler. With this background we can then dive more deeply into the compilation method in Sec.~\ref{sec:compilation} to show how we solve the second objective.

\subsection{Overview of Workflow}
\label{sec:nxtf_overview}
The two-step workflow to deploy a \ac{DNN} on Loihi using the NxTF interface is outlined in Fig.~\ref{fig:nxtf_workflow}. In the first step, the user trains or converts an \ac{SNN}, for instance using a third-party tool like Nengo \citep{Bekolay2014-zm}, \acs{SLAYER} \citep{Shrestha2018-io}, or the \acf{SNN TB} \citep{Rueckauer2017-ch}. The expected output of this step is a set of learned weights for each layer, the layer specifications defining \eg the padding and stride of convolutions, and the desired neuron properties (like voltage decay and threshold value). 

\begin{figure}[htb]
    \centering
    \includegraphics[width=\linewidth]{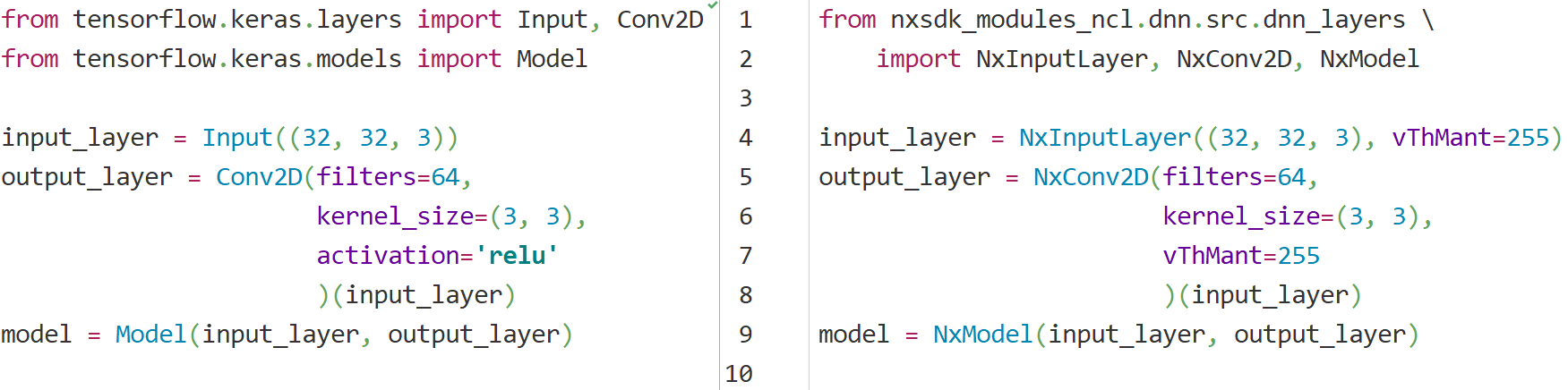}
    \caption{Comparison of model construction using \texttt{tf.keras} (left) and NxTF (right). The main difference is that class names get an \texttt{Nx} prefix and accept additional Loihi arguments (\eg for the voltage threshold in this example).}
    \label{fig:nxtf_codesample}
\end{figure}

Given this information, the user can then define a network in Python using the NxTF interface (red box in Fig.~\ref{fig:nxtf_workflow}). This interface is derived from Keras and follows the same syntax for setting up a Keras model, except that additional Loihi arguments (for threshold, time constants etc.) are supported. See Fig.~\ref{fig:nxtf_codesample} for a minimal code sample. The result at this stage is an instance of NxModel, which inherits all functionality of its Keras base class.

To prepare the NxModel for deployment on Loihi, the user then calls the \texttt{compile} method, which consists of two parts: A partitioner and a mapper. The objective of the partitioner is to find the optimal distribution of the network across neurocores, whereas the mapper is responsible for bit-level configuration of Loihi registers.

Once the mapping is complete, the user is able to run the model and retrieve the output for evaluation. A detailed tutorial covering these steps is provided within the \ac{INRC} framework\footnote{\url{https://github.com/intel-nrc-ecosystem/models}, accessed \today}. In Sec.~\ref{sec:nxtf_results}, we apply this pipeline on models trained directly on spike-based datasets as well as on models converted from frame-based \acp{ANN}. But before going to the results we will take a closer look at the Loihi hardware and NxTF compiler. 

\subsection{Loihi}
\label{sec:loihi}
\subsubsection{Architecture Overview}
Intel's Loihi research chip is an asynchronous, compute-in-memory neuromorphic processor optimized for the execution of \aclp{SNN}. Fabricated in Intel's standard 14 nm CMOS process, Loihi consists of 128 neurocores, each of which supports up to 1024 neurons. Three embedded x86 processors per chip enable off-chip data encoding and interaction with the neurocores. Loihi's asynchronous network-on-chip communication infrastructure allows it to be seamlessly scaled up to various form factors, ranging from 2-chip USB device \textit{Kapoho Bay} to the 768-chip rack \textit{Pohoiki Springs} \citep{Frady2020-lj}. 

Spiking neurons on Loihi are stateful dynamical systems that support a wide range of features. These include synaptic plasticity, variable numeric precision of synaptic weights up to 9 bits, multi-compartment models, threshold adaptation for homeostasis, and configurable synaptic, axon and refractory time constants. To support hierarchical and repeated connectivity as in \acp{CNN}, Loihi provides a connection-sharing mechanism (described in more detail in Sec.~\ref{sec:compilation}).

The dynamical equations that underlie the Loihi neuron model are approximated in discrete time. Unlike in conventional synchronous architectures, the transition between algorithmic time steps is not driven by a fixed global clock but mediated through a barrier synchronization protocol between all participating neurocores. As part of this protocol, neurocores signal the completion of the workload for the current time step to their neighbors, resulting in a workload-dependent duration of each time step. Neuron updates contribute on the order of microseconds to the duration of each time step. Spike traffic typically dominates the overall time step duration. Hence, spatially and temporally sparsely communicating \acp{SNN} promise the greatest level of efficiency on such architectures \citep{Mostafa2017-xr,Rueckauer2018-sz}. 

\begin{figure}[htb]
    \centering
    \includegraphics[width=0.7\linewidth]{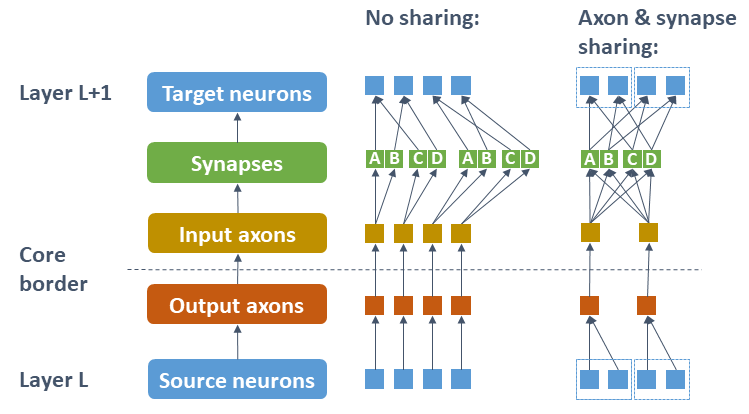}
    \caption{Fundamental components of a neurocore in Loihi (left column) and information flow in the absence (center) and presence (right) of resource sharing. Dashed blue boxes denote populations of neurons that make use of shared axons and synapses. 
    }
    \label{fig:neurocore_pipeline}
\end{figure}

\subsubsection{Information Flow in Loihi Cores}
In order to understand how the compiler utilizes Loihi's resource sharing features to efficiently compress \acp{CNN}, we will briefly review the logical entities within a neurocore (\cf left column of Fig.~\ref{fig:neurocore_pipeline}). 
The central building blocks are compartments, which may form single- or multi-compartment neurons. For simplicity we will use the term "neuron" in this section to denote a single compartment.\footnote{A two-compartment neuron will be introduced in Sec.~\ref{sec:soft_reset}.} A neuron accumulates spikes and eventually generates a spike itself once its membrane voltage exceeds a threshold. Spikes are routed from a neuron to their destination via axons. In our compiler, we distinguish between output and input axons. An output axon is responsible for core-to-core connectivity; it routes spikes to a corresponding input axon on the same or another core. The input axon references a list of synapses, which connect the axon to its destination neurons. Spikes travelling through the axon are weighted by the respective synaptic strength before accumulation on the neuron's membrane potential. While this level of abstraction is sufficient for the purpose of the present work, a more detailed description can be found in \cite{Lin2018-fg}.

\begin{table}[htb]
\caption{Rough estimate of resource constraints on a Loihi chip.}
\label{tab:core_constraints}
\centering
\begin{tabular}{l c l} 
 \toprule
 Resource & Availability & Assumptions \\
 \midrule
 Cores & 128 / chip & - \\
 Neurons & 1024 / core & Single compartment \\
 Input axons & 4096 / core & No off-chip axons \\
 Output axons & 4096 / core & No off-chip axons \\
 \multirow{2}{*}{Synapses} & \multirow{2}{*}{128k / core} & 8 bit weights; \\
  & & depends on compression \\
\end{tabular}
\end{table}

The quantity of these entities (neurons, axons, synapses) on a given neurocore is limited by the \acs{SRAM} memory resources in each neurocore (\cf Table \ref{tab:core_constraints}). Fortunately, Loihi provides a way to share axonal and synaptic resources to efficiently compress highly redundant convolutional connectivity patterns (Fig.~\ref{fig:neurocore_pipeline} right). 

In particular, all neurons in a source population of neurons (dashed box in Fig.\ref{fig:neurocore_pipeline}) may share one output axon to connect to a shared input axon instead of consuming discrete axonal resources for every neuron in the population. Similarly, these special shared input axons, may reference a shared set of synapses (A, B, C, D) rather than redundant copies of those synapses. Exploiting this resource sharing while adhering to core constraints is the aim of the NxTF compiler, which we address next.

\subsection{Compilation Method}
\label{sec:compilation}
Given a network with certain topology, the purpose of the compiler is to find a distribution of neurons across multiple neurocores that makes optimal use of the available chip resources. Due to the layer-to-layer connectivity, the resource requirements of a layer depend on the way the subsequent layer is distributed across neurocores. Therefore layers cannot be partitioned independently of each other. On the other hand, partitioning all layers simultaneously is unfeasible because the combinatorial space of possible partitions is large. 

Instead, we opt for a greedy two-layer-wise optimization, which will be explained in Sec.~\ref{sec:optimization}. 
To quantify the degree of optimality, we define a cost function (Sec.~\ref{sec:cost_function}) that takes into account hard constraints (\eg number of available neurons per core) as well as soft constraints (\eg the total number of cores used.). To understand the terms contributing to the total cost, we need to compare opportunities for resource sharing in \acsp{DNN} and on Loihi (Sections \ref{sec:connection_sharing} and \ref{sec:axon_sharing}).

\begin{figure}[htb]
    \centering
    \includegraphics[width=0.7\linewidth]{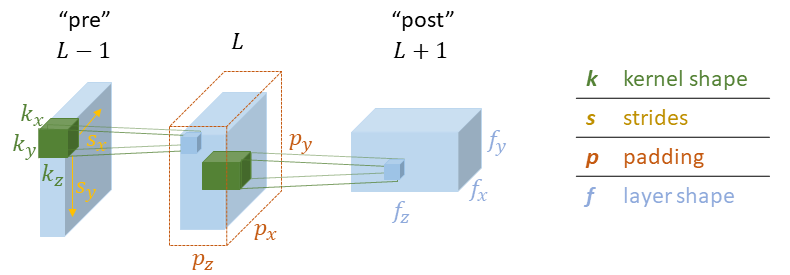}
    \caption{Connectivity pattern of a \ac{CNN}. One set of kernels (green block) is shared among neurons of a given layer, and applied by striding (yellow arrows) along the height and width of the feature map (blue box). The layer may optionally be padded with zeros to maintain the same dimensions after convolution.}
    \label{fig:cnn_terminology}
\end{figure}

\subsubsection{Connection Sharing}
\label{sec:connection_sharing}
\aclp{DNN} are characterized by a regular connectivity and repeated architectural patterns. This fact can be exploited to develop a compiler optimized specifically for \acs{DL} applications. Commonly seen \textit{convolution} layers are particularly advantageous because of their efficient use of trainable parameters. For a given layer, a single set of weight kernels is re-used at every spatial location of the feature map. Fig.~\ref{fig:cnn_terminology} illustrates this connectivity pattern within a \ac{CNN} and defines basic terminology used throughout this section.

\begin{figure}[htb]
    \centering
    \includegraphics[width=0.7\linewidth]{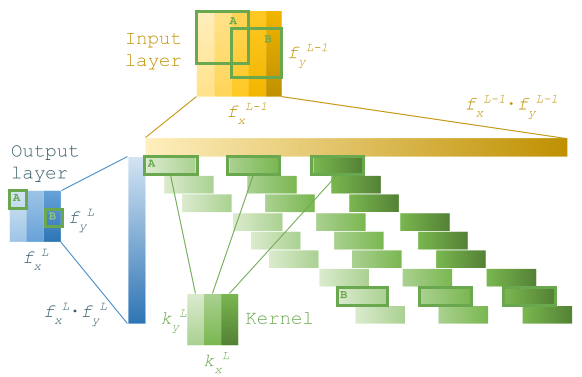}
    \caption{Unrolled connectivity matrix from an input layer (yellow) to an output layer (blue), via a convolution kernel (green). Boxes labeled A and B illustrate the application of the convolution kernel at two different locations of the feature map. }
    \label{fig:kernel_id_map}
\end{figure}

NxTF flattens all layers to a 1D topology on Loihi, so we introduce another way of inspecting the \ac{CNN} structure in Fig.~\ref{fig:kernel_id_map}. For clarity we consider only the spatial ($x,y$) dimensions of a feature map, neglecting the channel ($z$) dimension. The graphic illustrates the implied connectivity matrix resulting from the convolutional connection from the input layer (yellow) to the output layer (blue) with the layers flattened to 1D. This \textit{Toeplitz matrix} is well known in the context of unrolling convolutions as matrix-vector product \citep{Chetlur2014-cf}. Layers and kernels are color-coded by neuron and weight index, respectively. The two green boxes (labeled A, B) show two applications of the kernel at different locations of the feature map.

\begin{figure}[htb]
\centering
    \begin{subfigure}{.32\textwidth}
        \centering
        \includegraphics[width=0.99\linewidth]{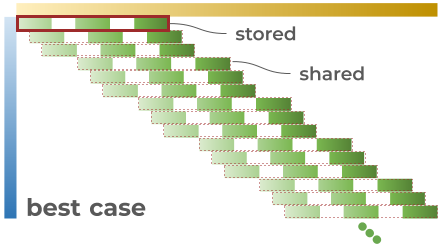}
        \caption{}
        \label{fig:connection_sharing_full}
    \end{subfigure} %
        \begin{subfigure}{.32\textwidth}
        \centering
        \includegraphics[width=0.99\linewidth]{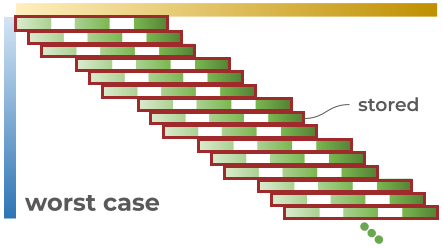}
        \caption{}
        \label{fig:connection_sharing_off}
    \end{subfigure} %
        \begin{subfigure}{.32\textwidth}
        \centering
        \includegraphics[width=0.99\linewidth]{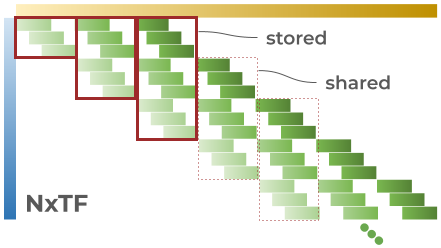}
        \caption{}
        \label{fig:connection_sharing_nxtf}
    \end{subfigure}
\caption{Optimal connection sharing in \acp{CNN} (\ref{fig:connection_sharing_full}), without weight re-use (\ref{fig:connection_sharing_off}), and with NxTF exploiting Loihi's resource sharing features (\ref{fig:connection_sharing_nxtf}).}
\label{fig:connection_sharing}
\end{figure}

With this perspective of the connectivity matrix at hand, we can compare weight sharing in a \ac{CNN} on general-purpose hardware against weight sharing on Loihi (Fig.~\ref{fig:connection_sharing}). Panel \ref{fig:connection_sharing_full} shows that, in a \ac{CNN}, we would ideally store the kernel only once, and re-use it when iterating over the output neurons (rows). However, in Loihi, weight look-up can only be triggered by the arrival of input spikes resulting in a column-major representation of the connectivity matrix. Hence, as described in Fig.~\ref{fig:neurocore_pipeline}, the synapses referenced by input axons form the non-zero entries in each column of Fig.~\ref{fig:connection_sharing_off}. Such a column vector does not appear with the same regularity as a row vector. 

Although Loihi does not support the minimal row-major representation of a convolutional kernel, the axon and synapse sharing mechanism described in Fig.~\ref{fig:neurocore_pipeline} still allows to exploit regularities in the column-major representation to efficiently compress large-scale \acp{DNN} such as MobileNet architectures as will be shown in Sec.~\ref{sec:nxtf_evaluation}. The basic principle is illustrated in Fig.~\ref{fig:connection_sharing_nxtf}. A population of input neurons shares a group of synapses (solid red rectangle), which is stored on chip. Whenever one of these synapse groups appears at other spatial locations, the synapses can be re-used by those other neuron populations as well (dotted red rectangles). 

The degree to which we can exploit this connection sharing depends on the particular way that the layer is partitioned across neurocores. Connection sharing is factored into the partitioner's cost function and thus is subject to the optimization procedure discussed below.

\subsubsection{Axon Sharing} 
\label{sec:axon_sharing}
Aside from neuron count and synaptic memory, another limited resource per neurocore is the number of core-to-core routing slots, the axons. In Fig.~\ref{fig:neurocore_pipeline} (right panel) we have shown how a population of source neurons (dashed blue box) shares the same axon. Likewise, we can define a population of input neurons in Fig.~\ref{fig:connection_sharing} (right panel), to contain all neurons of the input layer that utilize a particular synapse group (red box). This neuron population may use a single shared axon to send spikes to the output layer, where the spikes will be fanned out according to the synapse group. Without the ability to share axons, the layer would possibly have to be distributed across more cores to satisfy the axon constraint (Table \ref{tab:core_constraints}). 

\begin{figure}[htb]
    \centering
    \includegraphics[width=0.5\linewidth]{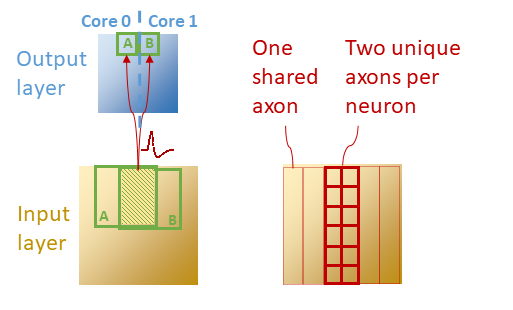}
    \caption{Axons in NxTF may be shared, except when targeting multiple cores, as in the example on the left. Following the notation of Fig.~\ref{fig:kernel_id_map}, the green boxes labeled A and B indicate the consecutive application of a convolution kernel. A spike originating from the shaded region of kernel overlap is sent to neurons on different cores in the output layer. Such connections require unique rather than shared axons (right panel).}
    \label{fig:axon_sharing}
\end{figure}

However, such axon sharing is possible only when spikes are being routed from one core to a single other core. If multiple cores are targeted, a discrete axon for each core is required. Fig.~\ref{fig:axon_sharing} illustrates this case in a convolution layer. The subsequent application of a convolution kernel in two neighboring locations of the input layer defines a region of kernel overlap (shaded box). Neurons within this region send their spikes to two neighboring neurons ($A, B$) in the output layer. If the layer happens to be partitioned such that neuron $A$ lives in core 0 but neuron $B$ in core 1, the input neurons within the shaded region each need two discrete axons to reach $A$ and $B$ on their separate cores. The input neurons outside the kernel overlap may share their axons as indicated on the right side of Fig.~\ref{fig:axon_sharing}. As with connection sharing, the amount of shared axons is included in the cost function and optimization algorithm. 

\begin{figure}[htb]
    \centering
    \includegraphics[width=0.4\linewidth]{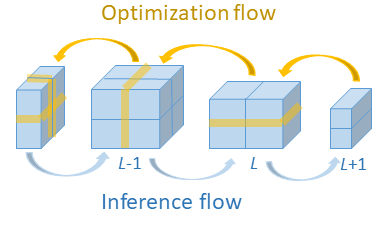}
    \caption{The way layer $L+1$ is partitioned (blue line denotes core border) effects a duplication of axons in layer $L$ for neurons connecting to multiple cores in layer $L+1$ (thick yellow line), thus determining the reverse optimization direction.}
    \label{fig:multiplicity}
\end{figure}

\subsubsection{Optimization procedure} 
\label{sec:optimization}
We saw in Fig.~\ref{fig:axon_sharing} that a given partitioning of layer $L$ influences the resource allocation in the next lower ("pre-") layer $L-1$, namely the axon count in the depicted case. This circumstance implies that our optimization algorithm needs to traverse the network graph from top to bottom: A given layer can only be partitioned once we know the resource duplication imposed by its post-layer (Fig.~\ref{fig:multiplicity}).

\begin{figure}[htb]
    \centering
    \includegraphics[width=\linewidth]{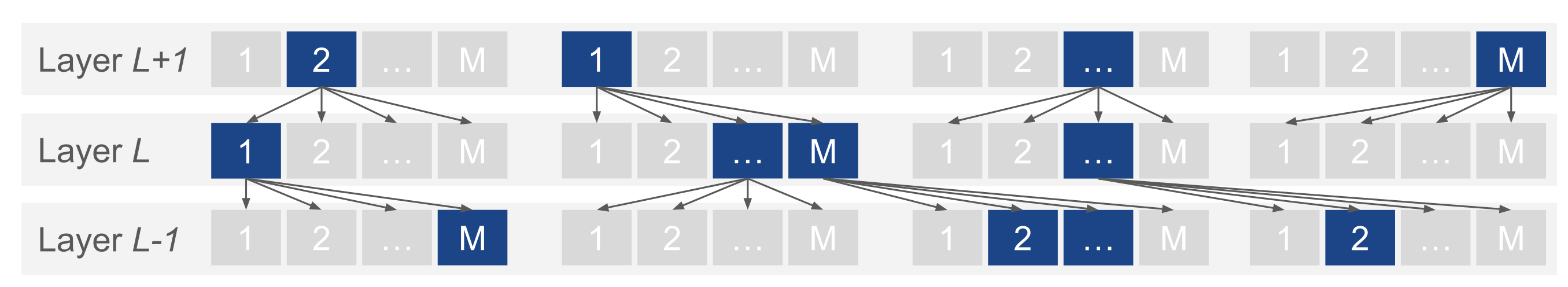}
    \caption{Each layer selects $M$ possible partitions (blue) among $M^2$ candidates according to cost function \eqref{eq:cost}.}
    \label{fig:partition_candidates}
\end{figure}

To optimize the distribution of a layer $L$, the partitioner proposes $M$ different \textit{partition candidates} and computes their cost according to Eq.~\eqref{eq:cost} (Sec.~\ref{sec:cost_function}). But since layers cannot be optimized in isolation, the partitioner moves on to layer $L-1$ and proposes another $M$ partition candidates for each of the $M$ candidates of layer $L$. If we were to continue like this for a simultaneous optimization of the whole network, the combinatorial space would quickly become intractable because of the exponential growth in the number of partition candidates. Instead, we settle on a greedy optimization approach where the number of candidates per layer never exceeds $M^2$. After proposing $M^2$ candidates for layer $L-1$ ($M$ for each of the $M$ candidates of layer $L$), the partitioner selects the $M$ best ones according to the combined cost of layer $L-1$ and all its parent layers (index $\ge L$). Fig.~\ref{fig:partition_candidates} illustrates this selection process.

\subsubsection{Cost Function} 
\label{sec:cost_function}
The efficiency of a given layer partition is evaluated based on hard and soft constraints. Violation of a hard constraint (like exceeding the number of neurons per core) immediately invalidates a partition candidate. These limits are defined by available hardware resources on neurocores and are listed in Table \ref{tab:core_constraints}. Soft constraints concern the flexible allocation of resources within the hard limits, and count towards the total cost of a partition within the optimization algorithm. For instance, distributing a network of 10 neurons across 10 cores ensures the hard constraints but is suboptimal if the neurons could fit on a single core.

The total cost of layer $L$ with respect to soft constraints can be written as
\begin{equation}
    \mathcal{C}_\mathrm{soft}^L = \alpha_0 \cdot N_\mathrm{cores}^L + \alpha_1 \cdot \frac{N_\mathrm{syn}^L}{N_\mathrm{syn}^\mathrm{max}} + \alpha_2 \cdot \frac{N_\mathrm{axons}^L}{N_\mathrm{axons}^\mathrm{max}} + \alpha_3 \cdot N_\mathrm{offchip}^L,
\label{eq:cost}
\end{equation}
\noindent
where the first term counts the number of cores, the second term the synaptic resources, the third the number of axons, and the last accounts for the cost of routing spikes from one chip to another (which costs twice the number of axons as for on-chip core-to-core connections). The synapse and axon cost terms are normalized per core for compatibility with the first term. The $\alpha_i$-coefficients allow custom weighting of different cost terms but are uniformly set to 1 in our experiments.

\subsubsection{Synapse Compression}
\label{sec:synapse_compression}
In order to compactly encode synaptic lists, Loihi supports several compression methods: \texttt{sparse}, \texttt{dense}, and \texttt{run-length}. Currently, this option is set by the user (or left at default). A future version of NxTF could determine the optimal compression scheme at compile-time.


\subsubsection{Soft Reset}
\label{sec:soft_reset}
A common way of resetting the membrane potential after a spike is to set it to zero. \cite{Rueckauer2017-ch} have shown that in a time-stepped simulation there is some information loss associated with this \textit{hard reset}, which leads to an increased classification error. One alternative reset mechanism reduces the membrane potential by the threshold magnitude, which retains any excess charge and prevents information loss. 

To enable this \textit{soft reset} mode in NxTF, we exploited Loihi's support for multi-compartment neurons. In \textit{hard reset} mode, our cells are point neurons; in \textit{soft reset}, they consist of two compartments, where one fulfills the role of the soma and accumulates voltage to generate spikes. The second compartment is recurrently connected to the first and becomes active only when the soma crosses threshold and issues a spike. The second compartment then inhibits the soma via the recurrent connection with an amount of charge equal to the threshold.

The optional use of this \textit{soft reset} implementation improves error rate and reduces runtime, at the cost of doubling the number of compartments allocated for the network.

\subsubsection{Parameter Normalization}
\label{sec:nxtf_normalization}
When porting floating-point weights from an \ac{ANN} to Loihi, two issues need to be considered. First, Loihi expects fixed-point weight, bias and threshold values. Second, previous work has shown that the limited dynamic range of \acp{SNN} can lead to saturating or vanishing firing rates unless the network parameters are properly rescaled \citep{Diehl2015-vu,Rueckauer2017-ch}. The latter only concerns models converted from \acp{ANN}, not directly trained \acp{SNN}. To address these two points, we briefly summarize a normalization algorithm that ensures that parameters satisfy the hardware requirements, and that neuron activity spans the full dynamic range.

\paragraph{Integer conversion.}
For the weights $W$ of each layer we apply the following steps: 
\begin{enumerate}
    \item Determine the maximum or some high percentile of the weight distribution: $\sigma = \mathrm{max}(W)$. 
    \item Normalize values in $W$ by $\sigma$.
    \item Scale by the highest value allowed for weights, \ie $\mathrm{max}(-W^-_\mathrm{lim}, W^+_\mathrm{lim})$. The weight limits $W^{\pm}_\mathrm{lim}$ depend on the number of bits allocated for the weights. 
    \item Truncate values to the nearest integer.
    \item Clip the resulting weight values to the allowed range $[W^-_\mathrm{lim}, W^+_\mathrm{lim}]$. Clipping is only necessary if $\sigma$ was computed based on a percentile rather than the max.
\end{enumerate}

Biases can in principle be converted in the same way. A subtle difference arises from the fact that the allowed bit precision differs for weights and biases. If one wishes to fully exploit their individual bit precision, determining the normalization and scaling factors is more involved; details can be found in the implementation.

\paragraph{Dynamic range.}
To obtain the scaling factor for the dynamic range, we iterate over each layer in turn and estimate the expected distribution of net voltage change $\frac{\mathrm{d}u}{\mathrm{d}t} = W x + b$ to neurons in the layer, using a subset of the training data. To estimate the expected input distribution without having to run the \ac{SNN}, we build an \ac{SNN} emulation from a modified copy of the \ac{ANN}. In this model we perform the same changes to the weights as we would do to the \ac{SNN} (\ie integer quantization as outlined above), and mimick the \ac{SNN} activation function by dividing by the threshold $\tau$, which converts the membrane potential change into a spike rate.

If the maximum (or some high percentile) $\lambda = \mathrm{max}(\frac{\mathrm{d}u}{\mathrm{d}t})$ of the input distribution lies near the voltage threshold $\tau$, then this layer is utilizing the full dynamic range, and no change is required. Conversely, a supra-threshold value $\lambda > \tau$ indicates saturation, because neurons cannot fire more than one spike per time step. Small net inputs $\lambda < \tau$ predict that a large fraction of neurons will have low or possibly vanishing firing rates.

To prevent vanishing or saturating spike rates, the measured scaling factor $\lambda$ is used to scale the weights and biases as described in \cite{Rueckauer2017-ch}. Then, the layer parameters are transformed to integers as described above. Finally, these values are decomposed into mantissa and exponent parts as expected by Loihi.


\section{Results}
\label{sec:nxtf_results}

Our experimental results focus on two areas: The evaluation of the compiler, and the demonstration of NxTF on standard use cases. To address the first, we characterize the NxTF compiler by comparing the resource utilization on neurocores when mapping different convolutional architectures. Further, we perform a scaling analysis on a 28-layer MobileNet \citep{Howard2017-yb} to determine the execution time, energy consumption, and the number of cores required to fit networks of increasing size on the hardware. This analysis also reveals how effective the compiler is at exploiting Loihi's support for weight sharing. 

For the second area, we apply the NxTF framework in two common use cases: A) Frame-based \acp{ANN} are trained on standard image datasets, converted to \acp{SNN} \citep{Rueckauer2017-ch}, and then evaluated on Loihi. B) \acp{SNN} are trained on event-based datasets using \acs{SLAYER} \citep{Shrestha2018-io}, and are then deployed on Loihi. In addition to reporting the classification error for the tested models, we perform a profiling of power consumption and execution time, and compare against 
results published for other neuromorphic platforms.

\begin{figure}[htb]
\centering
\includegraphics[width=0.8\linewidth]{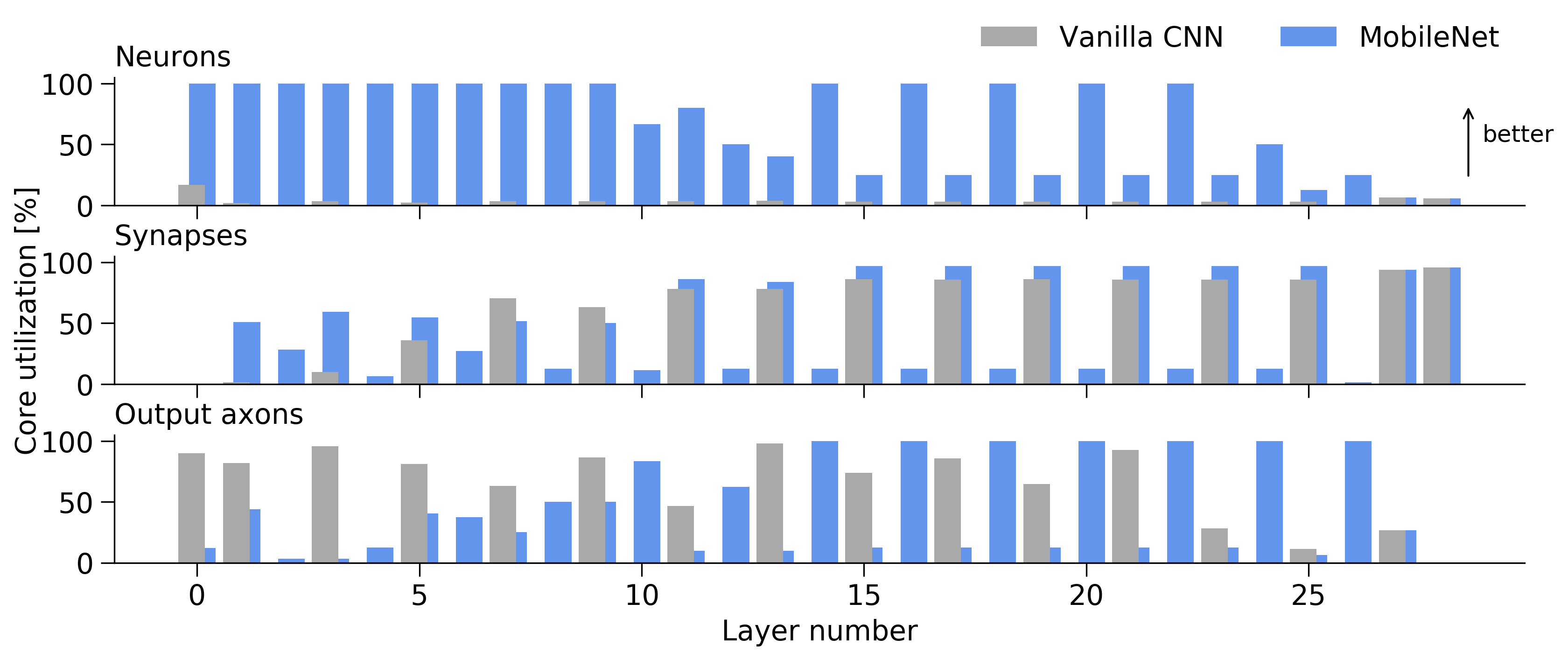}
\caption{Utilization of available neurons, synapses, and axons on neurocores in a standard \ac{CNN} (grey), and in a MobileNet architecture of equivalent size (blue).}
\label{fig:core_utilization}
\end{figure}

\subsection{NxTF Compiler Efficiency}
\label{sec:nxtf_evaluation}

\subsubsection{Depthwise Convolutions Maximize Core Utilization}
\label{sec:core_utilization}

We begin by showing a limitation with the current \ac{CNN} support on Loihi, and how to avert it. As discussed in Sec.~\ref{sec:axon_sharing}, axons can only be shared when targeting neurons within one core; if the synaptic fan-out spans across core borders (\cf Fig.~\ref{fig:axon_sharing}), neurons use a discrete axon for each destination core. A particularly adverse condition occurs when the target layer is split along the channel dimension: In that case, each neuron in the source layer sends its spikes to at least two separate cores, and axon sharing for that layer is no longer possible. This situation is suboptimal because a core will run out of available axons while using up only few of the other (in particular neuron) resources, resulting in an overall larger number of cores. We illustrate this case in Fig.~\ref{fig:core_utilization} for a \ac{CNN} of comparable size to MobileNet. The utilization of core resources is dominated by axons and synapses, while only a small fraction of available neurons are in use. 

Depthwise-separable convolutions provide a remedy to this imbalanced core utilization. Because depthwise convolutions operate on each channel separately rather than combining all channels, they reduce the overall synaptic fan-out. In addition, axons do not need to be duplicated even if the layer is split along the channel dimension, because the fan-out does not span the core borders. Such separable convolutions occur in MobileNet architectures \citep{Howard2017-yb}. By replacing the convolutions of the vanilla \ac{CNN} in Fig.~\ref{fig:core_utilization} with depthwise-separable convolutions, we obtain a core utilization where most of the layers fully exploit the available neurons, in addition to maintaining a high level of axon and synapse utilization. Even though the number of neurons in MobileNet is $1.6\times$ larger than in the vanilla \ac{CNN}, 
the total number of cores needed for MobileNet is $12.6\times$ smaller
because of the efficient resource utilization.

\subsubsection{NxTF Approaches CNN-like Weight Sharing on Loihi}
\label{sec:sharing_evaluation}

In this section we investigate to what extent the NxTF compiler is able to exploit Loihi's axon and synapse sharing features to implement convolutional architectures efficiently. As real-world workload we take an off-the-shelf MobileNet \citep{Howard2017-yb}, a common image classification architecture consisting of 28 layers. In a scaling analysis, we increase the input size from $32\times32$ to $176\times176$, thereby increasing the spatial dimensions of the feature maps and thus the overall model size. Each member of this model family is compiled and mapped to Loihi. With this experiment we address three questions: 1) How does the number of required neurocores scale with the model size? 2) To what extent are the resources on these cores used up? 3) How effectively does the compiler optimize weight re-use - ranging between the worst case (storing all copies of a convolution kernel) and the best case (storing the kernel only once). 

\begin{figure}[htb]
    \centering
    \includegraphics[width=0.8\linewidth]{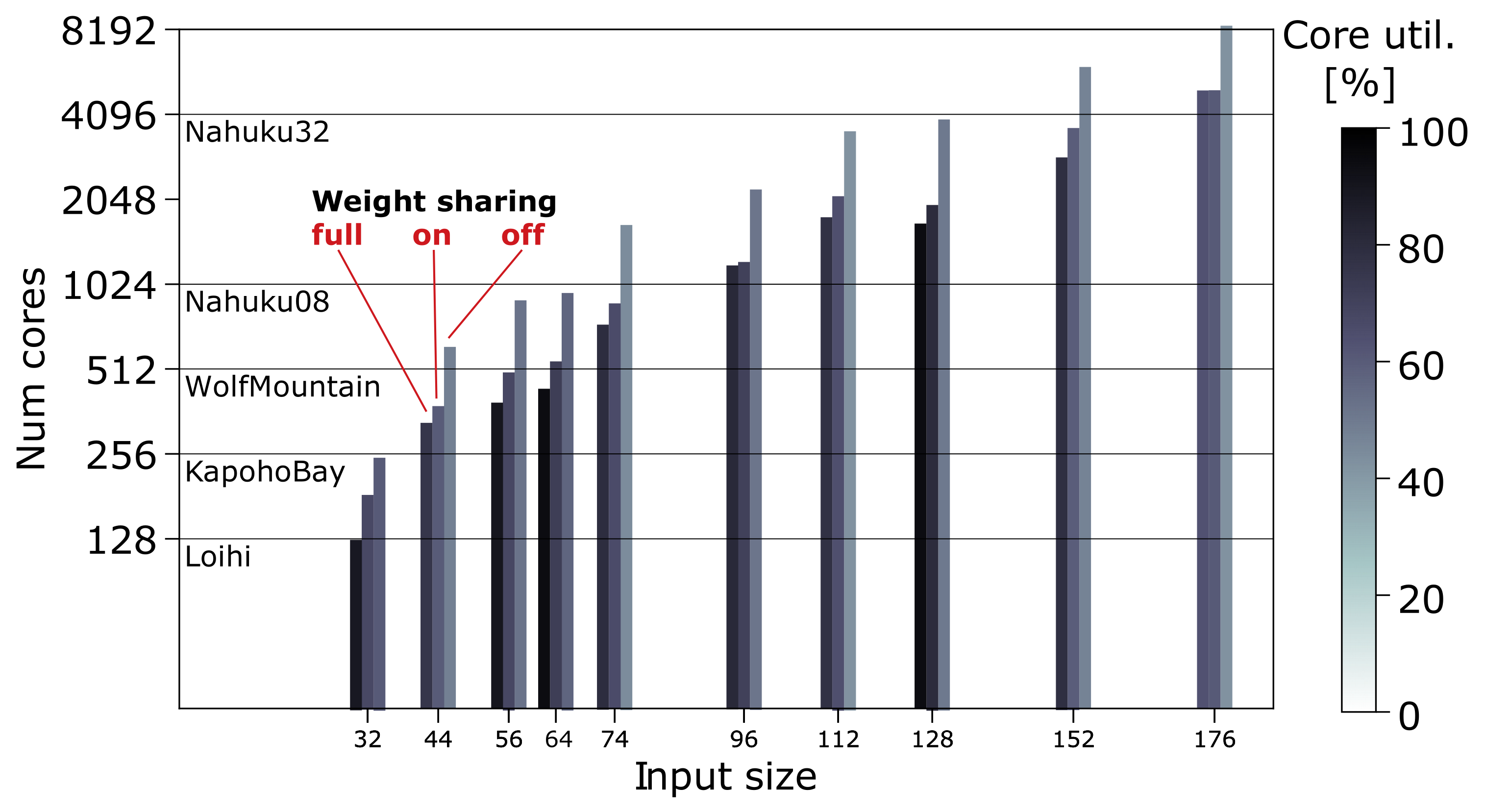}
    \caption{Scaling behavior of MobileNet. Each triplet of bars represents the compilation result for a given size of MobileNet, when weight sharing is turned \textit{off} (right), turned \textit{on} as supported by Loihi (center), and in the hypothetical best case of \textit{full} weight sharing (left). Bar height denotes the number of cores required to fit each compiled model. Bar shading indicates the amount of resource utilization per core (darker is better). The horizontal lines represent system form factors, ranging from a single Loihi chip to a 32 chip Nahuku board. The models use single-compartment neurons.}
    \label{fig:mobilenet_scaling}
\end{figure}

Fig.~\ref{fig:mobilenet_scaling} shows the number of cores used by MobileNet while increasing the model size. For a given model size, three bars are shown, corresponding to three repetitions of the compilation phase. The left bar of each triplet corresponds to the hypothetical best case scenario, in which the compiler computes the number of cores needed if each kernel had to be stored only once (Fig.~\ref{fig:connection_sharing_full}). The right bar of each triplet indicates the worst case, where weight sharing in NxTF is turned off and all kernel duplicates are stored (Fig.~\ref{fig:connection_sharing_off}). The middle bar shows the amount of resources allocated when weight sharing is turned on, \ie the NxTF compiler optimizes the layer partitions such as to maximize the amount of synapse re-use in Fig.~\ref{fig:connection_sharing_nxtf}.

From this comparison, we observe that the optimization for weight sharing as applied by NxTF reduces the number of cores by about $2\times$ on average, compared to when weight sharing is disabled. Furthermore, the partition solution found by the compiler comes close to the best case scenario, using only 13\% more cores on average. 

Another metric of interest is the core utilization, which we define as the average fraction of neurons used on the neurocores that are allocated by the compiler (gray shading in Fig.~\ref{fig:mobilenet_scaling}). This number should be as high as possible to minimize the overall number of cores, which impacts static power and to some degree active energy consumption. With weight sharing enabled, the solution found by the compiler deviates on average only 15\% from the hypothetical best case. This analysis demonstrates that despite the lack of dedicated support for convolutional compression, Loihi is able to encode such topologies close to the optimal encoding.

With this use of synapse sharing on Loihi, the compiler is able to fit a MobileNet of input size $152\times152$ onto the 32 chip Nahuku32 board at 60\% core utilization. The model with $128\times128$ input size fits on 16 chips at 80\% core utilization. This network consists of 28 layers, 5 M neurons, and 4 M parameters, thus representing a real-world workload for image classification on neuromorphic hardware. 


\subsection{SNN Performance Analysis}
\label{sec:nxtf_applications}
In this section, we use the NxTF framework to map directly trained \acp{SNN} as well as converted \acp{SNN} to Loihi and benchmark these networks on sparse event-based and dense frame-based datasets. Before presenting the results, we describe the methodology used to analyze \ac{SNN} performance on Loihi.

\subsubsection{Methodology}
\label{sec:snn_evaluation}

NxSDK allows for the measurement of both execution time per algorithmic time step and power consumption. These metrics allow us to compute the energy and time to solution per inference sample.

Algorithmic time steps in Loihi are further subdivided into sequential spiking, learning, and management phases, which are partially required for deadlock avoidance of the asynchronous mesh \citep{Davies2018-xo}. For inference-only \ac{DNN} workloads, only the spiking and management phase are of interest. During the spiking phase, neurons are updated and spikes are communicated between neurocores. During the management phase, the embedded \acp{CPU} may inject new data or read out the current \ac{SNN} state. When comparing our results to those of other groups, we compute the total execution time per algorithmic time step from the sum of all phases, and multiply by the number of time steps that each sample is run.


The total power consumption can be broken down into static and dynamic components. Static power mainly results from leakage while dynamic power results from active computations performed inside neurocores and embedded \acp{CPU}. Active neurocore power arises from memory access and logic operations for updating neurons and transmitting spikes. Active \ac{CPU} power arises from the clock of the synchronous x86 processors as well as the execution of user code. The power measurement utilities provided by NxSDK always measures the power and energy of the entire system but automatically excludes the part of unused neuro and \ac{CPU} cores. 

A common metric to combine both time and energy to solution is the \acf{EDP} \citep{Davies2019-xm}. It is computed from the product of energy and execution time. In our experiments, the energy per inference sample (frame) is computed from the product of execution time per sample and the total dynamic and static power of neurocores and x86 processors.\footnote{\label{footnote:specs}All measurements reported here are obtained using NxSDK version 0.9.5 on Nahuku32 board ncl-ghrd-04. \acs{CPU} results use an Intel Core i7-9700K processor with 32 GB \acs{RAM}. \acs{GPU} results use an Nvidia RTX 2070 card with 8 GB of memory. The operating system is Ubuntu 16.04.6 LTS running Python 3.5.5 and TensorFlow 1.13.1. Performance results are based on testing as of 1 Sep 2020 and may not reflect all publicly available security updates. Results may vary.}

\subsubsection{Models Trained with SLAYER on Event-Based Datasets}
\label{sec:slayer}

Our first use case for NxTF consists of deploying a model trained directly on spikes. We take as example the \acs{SLAYER} method \citep{Shrestha2018-io}. This spike-based training method uses a form of backpropagation that updates synaptic weights by taking into account the timing of spikes in the past. After constructing the model using the NxTF interface (\cf Fig.~\ref{fig:nxtf_codesample}), we transfer the publicly available \acs{SLAYER} weights and neuronal configuration from the original implementation\footnote{\url{https://github.com/bamsumit/slayerPytorch}, accessed Aug 2020} into our model, and invoke the NxTF compilation function. Neurons in \acs{SLAYER} models use a hard reset and thus require only a single compartment (\cf Sec.~\ref{sec:soft_reset}). The resulting network is then run on Loihi while measuring power consumption and execution time. A tutorial for porting a \acs{SLAYER}-trained model to Loihi using NxTF is included in the NxTF software package. 

\paragraph{N-MNIST}
\label{sec:nxtf_nmnist}
The N-MNIST dataset \citep{Orchard2015-fg} is an event-based version of the classic \acs{MNIST} handwritten digit dataset \citep{LeCun1989-mu}. The neuromorphic variant is obtained by recording the \acs{MNIST} digits using a \ac{DVS}, while performing saccades to elicit motion-induced events. The \acs{SLAYER} model for this dataset is a fully-connected network with a 1156-neuron input layer, a 512-neuron hidden layer, and a 10-neuron output layer. 

The network fits on 30 neurocores (about 25\% of a Loihi chip). Even though the number of neurons in this network is small, the all-to-all connectivity between layers leads to a large number of synapses in the hidden layer. Hence, the partitioner distributes this particular layer across 18 cores that are only partially occupied by neurons but whose synaptic memory is exploited to the limit. Similarly, the input layer is distributed across 11 cores, which do not exhaust the neuron capacity but rather the limit of available output axons (to accommodate the large fan-out to the hidden layer). The final layer (consisting of 10 neurons) fits on a single core, where it consumes less than 15\% of available neuron-, synapse- and axon-resources.


As in the original \acs{SLAYER} work, each of the 10 000 N-MNIST test samples is run for 350 algorithmic time steps, resulting in a total classification error of 1.49\% on Loihi. 
Table \ref{tab:summary} compares the error rate against a model trained on spikes using spatio-temporal backpropagation \citep{Patino-Saucedo2020-mx} and deployed on the SpiNNaker platform (unfortunately, power and execution time measurements were not available). 
With this \acs{SLAYER}-trained model, inference on N-MNIST can be done at 141 samples per second and with an energy consumption of 0.62 mJ per sample. 

As comparison, we trained the same architecture with standard backpropagation, using frames that were synthesized by binning 5 ms time slices from the \ac{DVS} event data. The resulting \ac{ANN} was then converted to an \ac{SNN} using the method by \cite{Rueckauer2017-ch}. Interestingly, the energy and time to solution is lower compared to the directly trained model. The reason is that with a denser frame input, the converted \ac{SNN} can be run for fewer algorithmic time steps to reach the desired classification error.

\paragraph{DVS Gestures}

To demonstrate that NxTF can also be applied to \acs{SLAYER}-trained \textit{convolutional} architectures, we repeated the procedure of N-\acs{MNIST} with the \ac{DVS} Gestures dataset, using the \acs{SLAYER} model from \cite{Shrestha2018-io}. This task consists of recognizing 11 human gestures like clapping and waving, recorded with a \ac{DVS}. The dataset was originally evaluated on TrueNorth \citep{Amir2017-pt}, to which we compare in Table \ref{tab:summary}. For consistency with their methodology, we adopt in this experiment their definition of "delay" or "time to solution", which is measured as the time it takes the correct neuron in the classification layer to fire its first spike after a new gesture is presented. For our model, this point is reached on average after 65 algorithmic time steps, resulting in an average delay of 4.35 ms per sample. Using instead a fix number of 300 time steps for measuring the delay, we obtain an execution time of 22 ms. The \acs{SLAYER} model consists of 6 layers, which the compiler fits on 59 neurocores - 14 cores less than a prior version implemented with NxNet \citep{Davies2020}.

Another Loihi implementation of this task was reported recently by \cite{Massa2020-wd}, who trained a model on synthesized frames, converted it using the method by \cite{Rueckauer2017-ch}, and applied NxTF to deploy the model on Loihi. The authors measure the delay based on 300 time steps per sample, which results in a 161 ms delay. 

\begin{table}[t]
\small
\caption{Summary and comparison of profiling results. "Energy" and "Delay" refer to energy and execution time per sample (\cf Sec.~\ref{sec:snn_evaluation}). The delay values denoted with a * in the \acs{DVS} Gestures section use the definition of delay from \cite{Amir2017-pt}, \ie the time it takes the correct neuron in the classification layer to become active. As for the Method column, "Train." denotes direct spike-based learning; "Conv." indicates a conversion approach; "\acs{ANN}" denotes running the original non-spiking model. 
Our work using NxTF is highlighted in bold. Remarks on other methods: \cite{Kiselev2016-pc} is an \acs{FPGA} implementation, \cite{Schmitt2017-cq} used 5 out of 10 classes and a subset of 1200 samples with resolution reduced to $10\times 10$.
}
\label{tab:summary}
\centering
\begin{tabular}{L{0.09\linewidth} L{0.11\linewidth} L{0.154\linewidth} R{0.065\linewidth} R{0.06\linewidth} R{0.06\linewidth} R{0.06\linewidth} R{0.06\linewidth} R{0.06\linewidth}} 
 \toprule
 Task & Platform & Method & Num Neurons & Num Params & Error [\%] & Energy [mJ] & Delay [ms] & \acs{EDP} [$\mu$Js] \\
 \midrule
 \hline
 \multirow{3}{*}{N-\acs{MNIST}} 
  & SpiNNaker & Train. (STBP) \cite{Patino-Saucedo2020-mx} & 810 & 792k & 2.08 & - & - & - \\
  & \textbf{Loihi} & \textbf{Train.} (\acs{SLAYER}) & 522 & 597k & 1.49 & 0.62 & 7.07 & 4.38 \\
  & \textbf{Loihi} & \textbf{Conv.} (\ac{SNN TB}) & 522 & 597k & 1.57 & 0.29 & 6.46 & 1.87 \\
 \hline
 \multirow{3}{0.9\linewidth}{\acs{DVS} Gestures} 
  & TrueNorth & Conv. (Corelets) \cite{Esser2016-ub} & 262k & 2M & 8.23 & 19 & 105* & 1971 \\
  & Loihi & Conv. (\ac{SNN TB}) \cite{Massa2020-wd} & 82k & 83k & 10.36 & - & 161 & - \\
  & \textbf{Loihi} & \textbf{Train.} (\acs{SLAYER}) & 31k & 1M & 3.79 & 0.54 & 4.35* & 2.34 \\
 \hline
 \multirow{9}{*}{\acs{MNIST}} 
  & Minitaur & Conv. (Matlab) \cite{Kiselev2016-pc} & 1k & 648k & 8 & - & 136 & - \\
  & SpiNNaker & Conv. (Matlab) \cite{Stromatias2015-kl} & 1k & 648k & 4.99 & 3.3 & 11 & 36.26 \\
  & TrueNorth & Conv. \cite{Esser2015-wd} & 4k & - & 0.60 & - & - & - \\
  & BrainScales & Conv. (PyNN) \cite{Schmitt2017-cq} & 55 & 2k & 5 & - & - & - \\
  & SpiNNaker & Conv. (\ac{SNN TB}) \cite{Patino-Saucedo2020-mx} & 8k & 213k & 1.80 & - & - & - \\
  & Loihi & Conv. (\ac{SNN TB}) \cite{Massa2020-wd} & 58k & 60k & 1.30 & - & 8 & - \\
  & \textbf{Loihi} & \textbf{Conv.} (\ac{SNN TB}) & 4k & 7k & 0.79 & 0.66 & 6.65 & 4.38 \\
  & \textbf{\acs{CPU}}\textsuperscript{\ref{footnote:specs}} & \textbf{\acs{ANN}} (Tensorflow) & 4k & 7k & 0.74 & 19 & 0.4 & 7.55 \\
  & \textbf{\acs{GPU}}\textsuperscript{\ref{footnote:specs}} & \textbf{\acs{ANN}} (Tensorflow) & 4k & 7k & 0.74 & 111 & 2 & 222 \\
 \hline
 \multirow{5}{*}{\acs{CIFAR}-10} 
  & TrueNorth & Conv. (Corelets) \cite{Esser2016-ub} & 1M & - & 17.50 & - & - & - \\
  & Loihi & Conv. (\ac{SNN TB}) \cite{Massa2020-wd} & 82k & 83k & 22.90 & - & 21 & - \\
  & \textbf{Loihi} & \textbf{Conv.} (\ac{SNN TB}) & 413k & 3M & 8.52 & 102 & 340 & 34926 \\
  & \textbf{\acs{CPU}}\textsuperscript{\ref{footnote:specs}} & \textbf{\acs{ANN}} (Tensorflow) & 413k & 3M & 8.07 & 157 & 3 & 508 \\
  & \textbf{\acs{GPU}}\textsuperscript{\ref{footnote:specs}} & \textbf{\acs{ANN}} (Tensorflow) & 413k & 3M & 8.07 & 1035 & 18 & 18924 \\
\end{tabular}
\end{table}

\subsubsection{Converted Models on Frame-Based Datasets}
\label{sec:nxtf_frame_based_results}

Our second use case for the NxTF framework covers the conversion of pre-trained \acp{ANN} to \acp{SNN}. For convenience we developed an interface\footnote{\url{https://github.com/intel-nrc-ecosystem/models/tree/master/nxsdk_modules_ncl/snntoolbox}} between NxTF and a common conversion software, the \acf{SNN TB} \citep{Rueckauer2017-ch}. This open-source framework automates most stages in the conversion and deployment pipeline. A user provides the pre-trained \ac{ANN} in one of the supported \ac{DL} libraries (Tensorflow / Keras, Pytorch, Caffe, Lasagne). The toolbox parses the provided network and constructs an equivalent model using the NxTF layer classes. If desired, \ac{SNN TB} performs a parameter normalization step as outlined in Sec.~\ref{sec:nxtf_normalization}. Subsequently, the network is compiled and run on Loihi. The toolbox includes methods for performance benchmarking and visualization of network activity.

\paragraph{MNIST}

For \acs{MNIST} we trained a 4-layer \ac{CNN} with Keras \citep{Chollet2015-lq} to achieve an error rate of 0.74\%. After conversion to an \ac{SNN} using the rate-based encoding method by \cite{Rueckauer2017-ch}, the \ac{SNN} was mapped to 14 neurocores on Loihi. Even though the model contains about $8 \times$ more neurons than the N-\acs{MNIST} model in Sec.~\ref{sec:nxtf_nmnist}, the core count is halved because the convolutional architecture requires about $88 \times$ fewer parameters than the fully-connected one. Thus, the utilization of resources on neurocores is dominated by the neuron count in case of the \ac{CNN}, rather than synapse / axon count as in the fully-connected network. The reduced core count also highlights the benefit of the basic connection sharing capabilities of Loihi as described in Sec.~\ref{sec:connection_sharing}. If no connections could be shared, the \ac{CNN} would require 341k discrete rather than 6746 shared weights, and the core count would likely approach that of the fully-connected model. 

After mapping the model onto Loihi, we ran it for 100 algorithmic time steps per sample. 
Table \ref{tab:summary} lists our profiling result together with related \acs{MNIST} results published on other hardware platforms using different network architectures. Unfortunately, energy and execution time were not always reported or could not be compared directly. For instance, the TrueNorth work reports a throughput of 1000 frames per second while consuming 0.1 mJ per frame. This throughput however is achieved by pipelining the processing of consecutive frames on individual layers of the network. Also, the first convolution layer is computed offline. Models on TrueNorth employ a customized convolution connectivity, are stateless and binarized, and neuron activations are encoded by single spikes rather than firing rates, which impedes a direct comparison.

\begin{figure}[htb]
    \centering
    \includegraphics[width=0.6\linewidth]{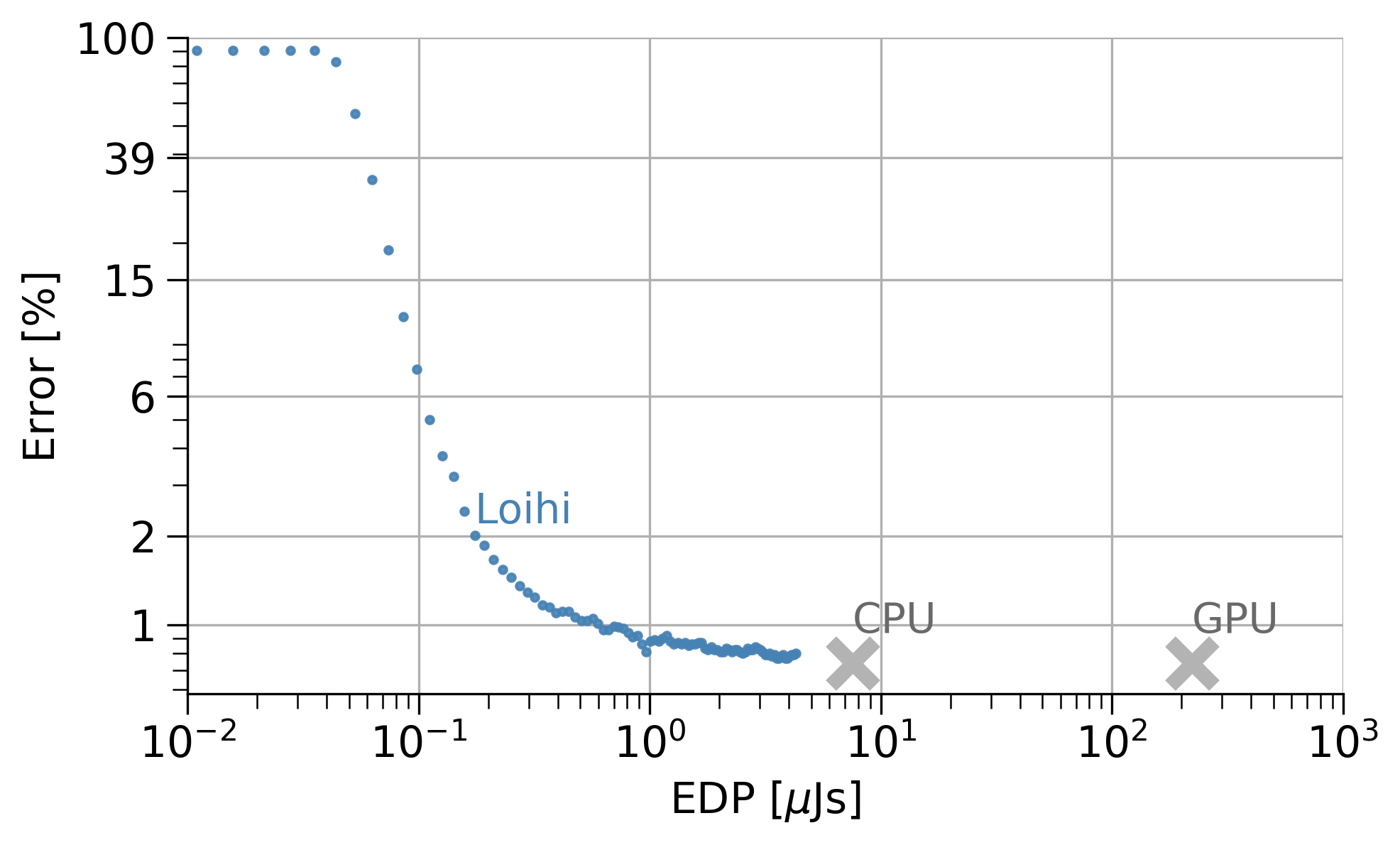}
    \caption{Trading off classification error against \ac{EDP} on the \acs{MNIST} task. Also shown are error and \ac{EDP} of the original \ac{ANN} before conversion, evaluated on a \acs{CPU} and \acs{GPU}.}
    \label{fig:err_vs_edp}
\end{figure}

A useful property of \acp{SNN} is their ability to trade off accuracy against computational cost. As the classification result is obtained by accumulating evidence in form of spikes at the output layer across a certain time period, one may shorten the run time of the network to reduce the inference delay and energy cost at the potential risk of misclassifying some samples. Fig.~\ref{fig:err_vs_edp} displays the resulting trade-off curve for the \acs{MNIST} model tested here. We observe a high error at the beginning of a simulation, where spikes have not yet reached the output layer. As spikes begin to accumulate at the output, the error drops and converges to the error rate achieved by the original \ac{ANN}, with an overall reduced \acl{EDP}. Note that the relatively low performance of the \acs{GPU} can be attributed to the fact that we target an online setting (\ie using a batch size of 1), which reduces the amount of parallelism that can be exploited.


\paragraph{CIFAR-10}

Based on the high core utilization observed in Sec.~\ref{sec:core_utilization} with this architecture, we choose MobileNet \citep{Howard2017-yb} for our experiments with \acs{CIFAR}-10. The model consists of 28 layers and fits on 861 cores when compiled by NxTF. We trained the network in Tensorflow to a classification error of 8.07\%, before converting it with \ac{SNN TB} and compiling it with NxTF. The model was then benchmarked while running for 400 time steps per frame. We also profiled the performance of the original \ac{ANN} on general-purpose hardware (Table \ref{tab:summary}) in online mode (batch size 1).

\begin{figure}[htb]
    \centering
    \includegraphics[width=0.5\linewidth]{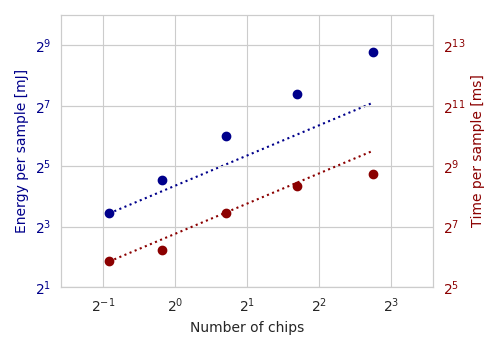}
    \caption{Loihi energy and execution time for MobileNet evaluated on \acs{CIFAR}-10, while increasing the model size by doubling the number of channels in each layer. The dashed lines denote linear scaling. Error bars are drawn but too small to be visible.}
    \label{fig:chip_scaling}
\end{figure}

While the model runs more energy-efficiently on Loihi than on the \acs{CPU} or \acs{GPU}, the execution time per frame is significantly larger than on those platforms. A different \acs{CIFAR}-10 model implemented with NxTF on Loihi by \cite{Massa2020-wd} runs $16\times$ faster than our MobileNet. This smaller model fits on only 37 cores, \ie 29\% of a single Loihi chip, compared to the 7 chips required by MobileNet. To investigate the dependence of the execution time on the model size, we performed a scaling analysis of the MobileNet architecture. MobileNet comes with a scale factor to reduce the number of channels in each feature map. By sweeping this scale factor from 1 to $1/16$ in steps of 2, we reduced the overall model size by the same factor. Each of the resulting models was mapped to Loihi and profiled for energy consumption and execution time per frame. Figure \ref{fig:chip_scaling} shows that the energy consumption increases supralinearly with the number of chips, while the execution time scales approximately linearly in this example. This delay scaling is in line with the observed increased run time of our larger MobileNet compared to the \ac{CNN} of \cite{Massa2020-wd}. 

The observed slowdown compared to \acs{CPU} or \acs{GPU} platforms for increasingly larger models can be attributed to three causes. 
1) Previous work on \acs{ANN}-to-\acs{SNN} conversion \citep{Rueckauer2017-ch} has shown that approximation errors in the converted model increase with the network depth. These errors can be reduced by increasing the number of algorithmic time steps, at the cost of an increased delay. 
2) The models evaluated here employ firing rates to represent and transmit information. The large number of spike events needed for an accurate encoding is reflected in a large number of synaptic operations, which slows down the execution duration of an algorithmic time step (Sec.~\ref{sec:snn_evaluation}). Promising approaches to reduce the spike count include temporal codes \citep{Mostafa2018-st,Rueckauer2018-sz}, spike frequency adaptation \citep{Zambrano2019-ru}, sigma-delta modulation \citep{Yousefzadeh2019-dd,Yu2020-xk}, and constrained training methods \citep{Sorbaro2020-dq,Rathi2020-lt,Patino-Saucedo2020-mx}. 
3) The large number of spikes present in a rate code have the added side effect of increasing the likelihood of spike congestion at Loihi's off-chip mesh links as discussed in \citet{Davies2020}. 
In our current implementation of the NxTF compiler, a neuron connected to multiple cores on another chip will send individual spikes to each off-chip destination. Especially in the case of increasing neuron fanout of large networks, this redundant spike traffic easily congests off-chip links and slows down execution significantly. To mitigate this congestion problem, the number of redundant spikes can be reduced either at the compiler or hardware level by sending only a single spike per neuron to each destination chip and then faning out to all destination cores from there. At the compiler level, this solution can be achieved by adding additional relay cores on the destination chip. At the hardware level, this multi-cast behavior could be integrated into the off-chip links.
Preliminary experiments at the compiler level (not shown here) suggest a potential speed-up of $15\times$ using relay cores.





\section{Discussion}

In this work we introduced NxTF, a programming interface and compiler embedded in Intel's NxSDK, for deploying \acp{DNN} on the neuromorphic Intel Loihi system. In this task we were guided by two objectives:
Reducing the entry barrier for users familiar with standard \ac{DL} software frameworks to access neuromorphic hardware, and optimal usage of the limited on-chip memory resources. The first objective was achieved by deriving the user-interface from Keras. The second objective was approached using a greedy layer-wise optimization algorithm that minimizes a cost metric as a function of neuron, synapse and axon requirements per Loihi core. Further, we exploited Loihi's support for connection sharing to minimize duplication of weight kernels in convolutional topologies.

The present work targets a particular neuromorphic platform, Loihi. Some aspects of NxTF are specifically tailored to this architecture, for instance the dissection of the connectivity matrix (Fig.~\ref{fig:connection_sharing_nxtf}) and the synapse encoding (Sec.~\ref{sec:synapse_compression}). However, several aspects of NxTF are general enough to be adopted in other \ac{SNN} hardware, such as the Keras-derived \ac{API} (Fig.~\ref{fig:nxtf_codesample}), or the partition optimization strategy (Sec.~\ref{sec:optimization}). 

A direct comparison against the compilation methods of other frameworks such as SpiNNTools \cite{Rowley2019-qf} or TrueNorth Corelets \citep{Amir2013-fr} is difficult because NxTF specializes in mapping \acp{DNN} rather than arbitrary network topologies. Perhaps most closely related is the Nengo tool chain \citep{Bekolay2014-zm} from Applied Brain Research, which provides a training and conversion framework for Deep Learning models (NengoDL) as well as a backend for deployment on Loihi. This framework has lead to promising results on a variety of tasks, including keyword spotting \citep{Blouw2019-el} and robotic control \citep{DeWolf2020-ey}. Unlike NxTF however, NengoDL currently does not optimize for synapse sharing in convolution layers and provides no automated partitioning scheme. To fit a large network onto Loihi, users must determine core shapes for a suitable layer dissection heuristically.

The NxTF framework was applied in two typical use cases: Models obtained via direct spike-based training and models converted from \acp{ANN} applied to sparse spike-based and dense frame-based inputs respectively. For the first use case we considered the neuromorphic datasets N-\acs{MNIST} and \acs{DVS} Gestures, for the second we converted a \ac{CNN} for \acs{MNIST} and a MobileNet for \acs{CIFAR}-10. 
The models achieve a throughput between 3 and 230 inference samples per second, consuming between 102 and 0.29 mJ per sample (for \acs{CIFAR}-10 and N-\acs{MNIST}, respectively). In terms of classification accuracy, our models are on par with other state-of-the-art models; with an error rate of 8.52\% on \acs{CIFAR}-10 we significantly advance the best result reported for this dataset on neuromorphic hardware.

Our results suggest several improvements for future neuromorphic architecture designs: Although the NxTF compiler achieves high resource utilization for many \ac{DNN} topologies as shown in Sec.~\ref{sec:nxtf_evaluation}, higher degrees of memory utilization would be possible via a shared-memory architecture that allowed to trade on-chip neuron, synapse or axon resources off against each other to ensure near 100\% memory utilization on every neurocore. This mechanism would avoid scenarios as shown in Fig.~\ref{fig:core_utilization} in which only a small fraction of neurons in a neurocore are occupied because the synaptic memory limits are exceeded. 
Support for dedicated \ac{CNN} compression would further allow to save on-chip memory resources.
Furthermore, a built-in soft reset as opposed to a hard reset mechanism would eliminate the need to duplicate neuron compartments as in our current multi-compartment implementation (Sec.~\ref{sec:soft_reset}). 
Finally, an execution slow-down due to spike congestion at off-chip mesh links could be mitigated by multi-cast links.

The present work also highlights the need for more efficient spike codes than the rate codes employed here. This need has long been recognized and a variety of alternatives have been proposed. Direct training of \acp{SNN} that tap into the temporal coding domain, explicit temporal coding schemes, hybrid training / conversion methods, or sigma-delta coding approaches are all promising steps in this direction \citep{Sorbaro2020-dq,Rathi2020-lt,Zambrano2019-ru,Yousefzadeh2019-dd,Yu2020-xk,Mostafa2018-st,Rueckauer2018-sz}.

Finally, performance results in the form of energy and execution time are scarce in the literature, even for well-explored datasets like \acs{MNIST}. We hope that the \ac{SNN} \ac{API} and compilation framework developed here will enable a more rapid and straightforward evaluation of novel \ac{SNN} models on neuromorphic hardware.

\begin{acks}
The majority of this work was performed during an internship of the two first authors at Intel Labs Oregon in 2019. In addition, this work was partially funded by the SNSF grant CRSII5 177255, the Samsung Advanced Institute of Technology, and the National Science Foundation Graduate Research Fellowship under Grant No. DGE 1752814.
\end{acks}

\begin{acronym}[SNN TB]
    \acro{ANN}{Artificial Neural Network}
    \acro{API}{Application Programming Interface}
    \acro{BN}{batch-normalization}
    \acro{CIFAR}{Canadian Institute for Advanced Research}
    \acro{CNN}{Convolutional Neural Network}
    \acro{CPU}{Central Processing Unit}
    \acro{DL}{Deep Learning}
    \acro{DNN}{Deep Neural Network}
    \acro{DRAM}{Dynamic Random Access Memory}
    \acro{DVS}{Dynamic Vision Sensor}
    \acro{EDP}{Energy Delay Product}
    \acro{Flop}{floating-point operation}
    \acro{FPGA}{field-programmable gate array}
    \acro{fps}{frames per second}
    \acro{GPU}{Graphics Processing Unit}
    \acro{IF}{Integrate-and-Fire}
    \acro{INRC}{Intel Neuromorphic Research Community}
    \acro{KLD}{Kullback-Leibler divergence}
    \acro{LSB}{least-significant bit}
    \acro{MAC}{multiply-accumulate}
    \acro{MNIST}{Modified National Institute of Standards and Technology}
    \acro{MSB}{most-significant bit}
    \acro{MSE}{mean square error}
    \acro{Op}{operation}
    \acro{PSP}{post-synaptic potential}
    \acro{RAM}{Random Access Memory}
    \acro{ReLU}{rectifying linear unit}
    \acro{RNN}{Recurrent Neural Network}
    \acro{SLAYER}{Spike Layer Error Reassignment}
    \acro{SNN}{Spiking Neural Network}
    \acro{SNN TB}{SNN toolbox}
    \acro{SRAM}{Static Random Access Memory}
    \acro{TPC}{temporal pattern code}
    \acro{TTFS}{time-to-first-spike}
    \acro{VGG}{Visual Geometry Group}
\end{acronym}

\bibliographystyle{ACM-Reference-Format}
\bibliography{paperpile, temp}


\begin{thebibliography}{41}


\ifx \showCODEN    \undefined \def \showCODEN     #1{\unskip}     \fi
\ifx \showDOI      \undefined \def \showDOI       #1{#1}\fi
\ifx \showISBNx    \undefined \def \showISBNx     #1{\unskip}     \fi
\ifx \showISBNxiii \undefined \def \showISBNxiii  #1{\unskip}     \fi
\ifx \showISSN     \undefined \def \showISSN      #1{\unskip}     \fi
\ifx \showLCCN     \undefined \def \showLCCN      #1{\unskip}     \fi
\ifx \shownote     \undefined \def \shownote      #1{#1}          \fi
\ifx \showarticletitle \undefined \def \showarticletitle #1{#1}   \fi
\ifx \showURL      \undefined \def \showURL       {\relax}        \fi
\providecommand\bibfield[2]{#2}
\providecommand\bibinfo[2]{#2}
\providecommand\natexlab[1]{#1}
\providecommand\showeprint[2][]{arXiv:#2}

\bibitem[\protect\citeauthoryear{Amir, Datta, Risk, Cassidy, Kusnitz, Esser,
  Andreopoulos, Wong, Flickner, Alvarez-Icaza, McQuinn, Shaw, Pass, and
  Modha}{Amir et~al\mbox{.}}{2013}]%
        {Amir2013-fr}
\bibfield{author}{\bibinfo{person}{A Amir}, \bibinfo{person}{P Datta},
  \bibinfo{person}{W~P Risk}, \bibinfo{person}{A~S Cassidy},
  \bibinfo{person}{J~A Kusnitz}, \bibinfo{person}{S~K Esser},
  \bibinfo{person}{A Andreopoulos}, \bibinfo{person}{T~M Wong},
  \bibinfo{person}{M Flickner}, \bibinfo{person}{R Alvarez-Icaza},
  \bibinfo{person}{E McQuinn}, \bibinfo{person}{B Shaw}, \bibinfo{person}{N
  Pass}, {and} \bibinfo{person}{D~S Modha}.} \bibinfo{year}{2013}\natexlab{}.
\newblock \showarticletitle{Cognitive Computing Programming Paradigm: A Corelet
  Language for Composing Networks of Neurosynaptic Cores}. In
  \bibinfo{booktitle}{\emph{The 2013 International Joint Conference on Neural
  Networks ({IJCNN})}}. \bibinfo{publisher}{IEEE}, \bibinfo{address}{Dallas,
  Texas, USA}, \bibinfo{pages}{1--10}.
\newblock


\bibitem[\protect\citeauthoryear{Amir, Taba, Berg, Melano, McKinstry, Di~Nolfo,
  Nayak, Andreopoulos, Garreau, Mendoza, Kusnitz, Debole, Esser, Delbruck,
  Flickner, and Modha}{Amir et~al\mbox{.}}{2017}]%
        {Amir2017-pt}
\bibfield{author}{\bibinfo{person}{A Amir}, \bibinfo{person}{B Taba},
  \bibinfo{person}{D Berg}, \bibinfo{person}{T Melano}, \bibinfo{person}{J
  McKinstry}, \bibinfo{person}{C Di~Nolfo}, \bibinfo{person}{T Nayak},
  \bibinfo{person}{A Andreopoulos}, \bibinfo{person}{G Garreau},
  \bibinfo{person}{M Mendoza}, \bibinfo{person}{J Kusnitz}, \bibinfo{person}{M
  Debole}, \bibinfo{person}{S Esser}, \bibinfo{person}{T Delbruck},
  \bibinfo{person}{M Flickner}, {and} \bibinfo{person}{D Modha}.}
  \bibinfo{year}{2017}\natexlab{}.
\newblock \showarticletitle{A Low Power, Fully {Event-Based} Gesture
  Recognition System}. In \bibinfo{booktitle}{\emph{2017 {IEEE} Conference on
  Computer Vision and Pattern Recognition ({CVPR})}}.
  \bibinfo{publisher}{IEEE}, \bibinfo{address}{Honolulu, HI, USA},
  \bibinfo{pages}{7388--7397}.
\newblock


\bibitem[\protect\citeauthoryear{Bekolay, Bergstra, Hunsberger, DeWolf,
  Stewart, Rasmussen, Choo, Voelker, and Eliasmith}{Bekolay
  et~al\mbox{.}}{2014}]%
        {Bekolay2014-zm}
\bibfield{author}{\bibinfo{person}{Trevor Bekolay}, \bibinfo{person}{James
  Bergstra}, \bibinfo{person}{Eric Hunsberger}, \bibinfo{person}{Travis
  DeWolf}, \bibinfo{person}{Terrence~C Stewart}, \bibinfo{person}{Daniel
  Rasmussen}, \bibinfo{person}{Xuan Choo}, \bibinfo{person}{Aaron~Russell
  Voelker}, {and} \bibinfo{person}{Chris Eliasmith}.}
  \bibinfo{year}{2014}\natexlab{}.
\newblock \showarticletitle{Nengo: A Python Tool for Building {Large-Scale}
  Functional Brain Models}.
\newblock \bibinfo{journal}{\emph{Front. Neuroinform.}} \bibinfo{volume}{7},
  \bibinfo{number}{JAN} (\bibinfo{year}{2014}), \bibinfo{pages}{1--13}.
\newblock


\bibitem[\protect\citeauthoryear{Blalock, Ortiz, Frankle, and Guttag}{Blalock
  et~al\mbox{.}}{2020}]%
        {Blalock2020-td}
\bibfield{author}{\bibinfo{person}{Davis Blalock}, \bibinfo{person}{Jose
  Javier~Gonzalez Ortiz}, \bibinfo{person}{Jonathan Frankle}, {and}
  \bibinfo{person}{John Guttag}.} \bibinfo{year}{2020}\natexlab{}.
\newblock \showarticletitle{What is the State of Neural Network Pruning?}. In
  \bibinfo{booktitle}{\emph{Proceedings of the 3rd {MLSys} Conference}}.
  \bibinfo{publisher}{MLSys}, \bibinfo{address}{Austin, TX, USA},
  \bibinfo{pages}{1--17}.
\newblock


\bibitem[\protect\citeauthoryear{Blouw, Choo, Hunsberger, and Eliasmith}{Blouw
  et~al\mbox{.}}{2019}]%
        {Blouw2019-el}
\bibfield{author}{\bibinfo{person}{Peter Blouw}, \bibinfo{person}{Xuan Choo},
  \bibinfo{person}{Eric Hunsberger}, {and} \bibinfo{person}{Chris Eliasmith}.}
  \bibinfo{year}{2019}\natexlab{}.
\newblock \showarticletitle{Benchmarking Keyword Spotting Efficiency on
  Neuromorphic Hardware}. In \bibinfo{booktitle}{\emph{Proceedings of the 7th
  Annual Neuro-inspired Computational Elements Workshop}} (Albany, NY, USA)
  \emph{(\bibinfo{series}{NICE '19}, \bibinfo{number}{Article 1})}.
  \bibinfo{publisher}{Association for Computing Machinery},
  \bibinfo{address}{New York, NY, USA}, \bibinfo{pages}{1--8}.
\newblock


\bibitem[\protect\citeauthoryear{Chetlur, Woolley, Vandermersch, Cohen, Tran,
  Catanzaro, and Shelhamer}{Chetlur et~al\mbox{.}}{2014}]%
        {Chetlur2014-cf}
\bibfield{author}{\bibinfo{person}{Sharan Chetlur}, \bibinfo{person}{Cliff
  Woolley}, \bibinfo{person}{Philippe Vandermersch}, \bibinfo{person}{Jonathan
  Cohen}, \bibinfo{person}{John Tran}, \bibinfo{person}{Bryan Catanzaro}, {and}
  \bibinfo{person}{Evan Shelhamer}.} \bibinfo{year}{2014}\natexlab{}.
\newblock \bibinfo{title}{{cuDNN}: Efficient Primitives for Deep Learning}.
  (\bibinfo{year}{2014}).
\newblock


\bibitem[\protect\citeauthoryear{Chollet}{Chollet}{2015}]%
        {Chollet2015-lq}
\bibfield{author}{\bibinfo{person}{Francois Chollet}.}
  \bibinfo{year}{2015}\natexlab{}.
\newblock \bibinfo{title}{Keras}.
\newblock
\newblock


\bibitem[\protect\citeauthoryear{Davidson, Br{\"u}derle, Eppler, Kremkow,
  Muller, Pecevski, Perrinet, and Yger}{Davidson et~al\mbox{.}}{2009}]%
        {Davidson2009-qh}
\bibfield{author}{\bibinfo{person}{Andrew~P Davidson}, \bibinfo{person}{Daniel
  Br{\"u}derle}, \bibinfo{person}{Jochen~M Eppler}, \bibinfo{person}{Jens
  Kremkow}, \bibinfo{person}{Eilif Muller}, \bibinfo{person}{Dejan Pecevski},
  \bibinfo{person}{Laurent Perrinet}, {and} \bibinfo{person}{Pierre Yger}.}
  \bibinfo{year}{2009}\natexlab{}.
\newblock \showarticletitle{{PyNN}: A Common Interface for Neuronal Network
  Simulators}.
\newblock \bibinfo{journal}{\emph{Front. Neuroinform.}} \bibinfo{volume}{2},
  \bibinfo{number}{11} (\bibinfo{year}{2009}), \bibinfo{pages}{1--20}.
\newblock


\bibitem[\protect\citeauthoryear{Davies}{Davies}{2019}]%
        {Davies2019-xm}
\bibfield{author}{\bibinfo{person}{Mike Davies}.}
  \bibinfo{year}{2019}\natexlab{}.
\newblock \showarticletitle{Benchmarks for Progress in Neuromorphic Computing}.
\newblock \bibinfo{journal}{\emph{Nature Machine Intelligence}}
  \bibinfo{volume}{1}, \bibinfo{number}{9} (\bibinfo{year}{2019}),
  \bibinfo{pages}{386--388}.
\newblock


\bibitem[\protect\citeauthoryear{Davies, Srinivasa, Lin, Chinya, Cao, Choday,
  Dimou, Joshi, Imam, Jain, Liao, Lin, Lines, Liu, Mathaikutty, McCoy, Paul,
  Tse, Venkataramanan, Weng, Wild, Yang, and Wang}{Davies
  et~al\mbox{.}}{2018}]%
        {Davies2018-xo}
\bibfield{author}{\bibinfo{person}{Mike Davies}, \bibinfo{person}{Narayan
  Srinivasa}, \bibinfo{person}{Tsung~Han Lin}, \bibinfo{person}{Gautham
  Chinya}, \bibinfo{person}{Yongqiang Cao}, \bibinfo{person}{Sri~Harsha
  Choday}, \bibinfo{person}{Georgios Dimou}, \bibinfo{person}{Prasad Joshi},
  \bibinfo{person}{Nabil Imam}, \bibinfo{person}{Shweta Jain},
  \bibinfo{person}{Yuyun Liao}, \bibinfo{person}{Chit~Kwan Lin},
  \bibinfo{person}{Andrew Lines}, \bibinfo{person}{Ruokun Liu},
  \bibinfo{person}{Deepak Mathaikutty}, \bibinfo{person}{Steven McCoy},
  \bibinfo{person}{Arnab Paul}, \bibinfo{person}{Jonathan Tse},
  \bibinfo{person}{Guruguhanathan Venkataramanan}, \bibinfo{person}{Yi~Hsin
  Weng}, \bibinfo{person}{Andreas Wild}, \bibinfo{person}{Yoonseok Yang}, {and}
  \bibinfo{person}{Hong Wang}.} \bibinfo{year}{2018}\natexlab{}.
\newblock \showarticletitle{Loihi: A Neuromorphic Manycore Processor with
  {On-Chip} Learning}.
\newblock \bibinfo{journal}{\emph{IEEE Micro}} \bibinfo{volume}{38},
  \bibinfo{number}{1} (\bibinfo{year}{2018}), \bibinfo{pages}{82--99}.
\newblock


\bibitem[\protect\citeauthoryear{Davies, Wild, Orchard, Sandamirskaya, Guerra,
  Joshi, Plank, and Risbud}{Davies et~al\mbox{.}}{2021}]%
        {Davies2020}
\bibfield{author}{\bibinfo{person}{Mike Davies}, \bibinfo{person}{Andreas
  Wild}, \bibinfo{person}{Garrick Orchard}, \bibinfo{person}{Yulia
  Sandamirskaya}, \bibinfo{person}{Gabriel A.~Fonseca Guerra},
  \bibinfo{person}{Prasad Joshi}, \bibinfo{person}{Philipp Plank}, {and}
  \bibinfo{person}{Sumedh Risbud}.} \bibinfo{year}{2021}\natexlab{}.
\newblock \showarticletitle{Advancing Neuromorphic Computing with Loihi: A
  Survey of Results and Outlook}.
\newblock \bibinfo{journal}{\emph{in review}}  \bibinfo{volume}{1}
  (\bibinfo{year}{2021}), \bibinfo{pages}{1}.
\newblock


\bibitem[\protect\citeauthoryear{DeWolf, Jaworski, and Eliasmith}{DeWolf
  et~al\mbox{.}}{2020}]%
        {DeWolf2020-ey}
\bibfield{author}{\bibinfo{person}{Travis DeWolf}, \bibinfo{person}{Pawel
  Jaworski}, {and} \bibinfo{person}{Chris Eliasmith}.}
  \bibinfo{year}{2020}\natexlab{}.
\newblock \showarticletitle{Nengo and {Low-Power} {AI} Hardware for Robust,
  Embedded Neurorobotics}.
\newblock \bibinfo{journal}{\emph{Front. Neurorobot.}}  \bibinfo{volume}{14}
  (\bibinfo{date}{Oct.} \bibinfo{year}{2020}), \bibinfo{pages}{568359}.
\newblock


\bibitem[\protect\citeauthoryear{Diehl and Cook}{Diehl and Cook}{2015}]%
        {Diehl2015-vu}
\bibfield{author}{\bibinfo{person}{Peter~U Diehl} {and}
  \bibinfo{person}{Matthew Cook}.} \bibinfo{year}{2015}\natexlab{}.
\newblock \showarticletitle{Unsupervised Learning of Digit Recognition Using
  {Spike-Timing-Dependent} Plasticity}.
\newblock \bibinfo{journal}{\emph{Front. Comput. Neurosci.}}
  \bibinfo{volume}{9}, \bibinfo{number}{August} (\bibinfo{date}{Aug.}
  \bibinfo{year}{2015}), \bibinfo{pages}{1}.
\newblock


\bibitem[\protect\citeauthoryear{Esser, Arthur, Merolla, Modha, and
  Appuswamy}{Esser et~al\mbox{.}}{2015}]%
        {Esser2015-wd}
\bibfield{author}{\bibinfo{person}{Steve~K Esser}, \bibinfo{person}{John~V
  Arthur}, \bibinfo{person}{Paul~A Merolla}, \bibinfo{person}{Dharmendra~S
  Modha}, {and} \bibinfo{person}{Rathinakumar Appuswamy}.}
  \bibinfo{year}{2015}\natexlab{}.
\newblock \showarticletitle{Backpropagation for {Energy-Efficient} Neuromorphic
  Computing}. In \bibinfo{booktitle}{\emph{Advances in Neural Information
  Processing Systems 28 ({NIPS} 2015)}}. \bibinfo{publisher}{Curran Associates,
  Inc.}, \bibinfo{address}{Montreal, Canada}, \bibinfo{pages}{1--9}.
\newblock


\bibitem[\protect\citeauthoryear{Esser, Merolla, Arthur, Cassidy, Appuswamy,
  Andreopoulos, Berg, McKinstry, Melano, Barch, di~Nolfo, Datta, Amir, Taba,
  Flickner, and Modha}{Esser et~al\mbox{.}}{2016}]%
        {Esser2016-ub}
\bibfield{author}{\bibinfo{person}{Steven~K Esser}, \bibinfo{person}{Paul~A
  Merolla}, \bibinfo{person}{John~V Arthur}, \bibinfo{person}{Andrew~S
  Cassidy}, \bibinfo{person}{Rathinakumar Appuswamy},
  \bibinfo{person}{Alexander Andreopoulos}, \bibinfo{person}{David~J Berg},
  \bibinfo{person}{Jeffrey~L McKinstry}, \bibinfo{person}{Timothy Melano},
  \bibinfo{person}{Davis~R Barch}, \bibinfo{person}{Carmelo di Nolfo},
  \bibinfo{person}{Pallab Datta}, \bibinfo{person}{Arnon Amir},
  \bibinfo{person}{Brian Taba}, \bibinfo{person}{Myron~D Flickner}, {and}
  \bibinfo{person}{Dharmendra~S Modha}.} \bibinfo{year}{2016}\natexlab{}.
\newblock \showarticletitle{Convolutional Networks for Fast, {Energy-Efficient}
  Neuromorphic Computing}.
\newblock \bibinfo{journal}{\emph{Proc. Natl. Acad. Sci. U. S. A.}}
  \bibinfo{volume}{113}, \bibinfo{number}{41} (\bibinfo{year}{2016}),
  \bibinfo{pages}{11441--11446}.
\newblock


\bibitem[\protect\citeauthoryear{Frady, Orchard, Florey, Imam, Liu, Mishra,
  Tse, Wild, Sommer, and Davies}{Frady et~al\mbox{.}}{2020}]%
        {Frady2020-lj}
\bibfield{author}{\bibinfo{person}{E~Paxon Frady}, \bibinfo{person}{Garrick
  Orchard}, \bibinfo{person}{David Florey}, \bibinfo{person}{Nabil Imam},
  \bibinfo{person}{Ruokun Liu}, \bibinfo{person}{Joyesh Mishra},
  \bibinfo{person}{Jonathan Tse}, \bibinfo{person}{Andreas Wild},
  \bibinfo{person}{Friedrich~T Sommer}, {and} \bibinfo{person}{Mike Davies}.}
  \bibinfo{year}{2020}\natexlab{}.
\newblock \showarticletitle{Neuromorphic Nearest Neighbor Search Using Intel's
  Pohoiki Springs}. In \bibinfo{booktitle}{\emph{{ACM} International Conference
  Proceeding Series}}. \bibinfo{publisher}{Association for Computing
  Machinery}, \bibinfo{address}{New York, NY, USA}, \bibinfo{pages}{1--9}.
\newblock


\bibitem[\protect\citeauthoryear{Furber, Lester, Plana, Garside, Painkras,
  Temple, and Brown}{Furber et~al\mbox{.}}{2013}]%
        {Furber2013-ws}
\bibfield{author}{\bibinfo{person}{S Furber}, \bibinfo{person}{D Lester},
  \bibinfo{person}{L Plana}, \bibinfo{person}{J Garside}, \bibinfo{person}{E
  Painkras}, \bibinfo{person}{S Temple}, {and} \bibinfo{person}{A Brown}.}
  \bibinfo{year}{2013}\natexlab{}.
\newblock \showarticletitle{Overview of the {SpiNNaker} System Architecture}.
\newblock \bibinfo{journal}{\emph{IEEE Trans. Comput.}} \bibinfo{volume}{62},
  \bibinfo{number}{12} (\bibinfo{year}{2013}), \bibinfo{pages}{2454--2467}.
\newblock


\bibitem[\protect\citeauthoryear{Howard, Zhu, Chen, Kalenichenko, Wang, Weyand,
  Andreetto, and Adam}{Howard et~al\mbox{.}}{2017}]%
        {Howard2017-yb}
\bibfield{author}{\bibinfo{person}{Andrew~G Howard}, \bibinfo{person}{Menglong
  Zhu}, \bibinfo{person}{Bo Chen}, \bibinfo{person}{Dmitry Kalenichenko},
  \bibinfo{person}{Weijun Wang}, \bibinfo{person}{Tobias Weyand},
  \bibinfo{person}{Marco Andreetto}, {and} \bibinfo{person}{Hartwig Adam}.}
  \bibinfo{year}{2017}\natexlab{}.
\newblock \showarticletitle{{MobileNets}: Efficient Convolutional Neural
  Networks for Mobile Vision Applications}.
\newblock \bibinfo{journal}{\emph{CoRR}}  \bibinfo{volume}{abs/1704.04861}
  (\bibinfo{year}{2017}), \bibinfo{pages}{1}.
\newblock


\bibitem[\protect\citeauthoryear{Kiselev, Neil, and Liu}{Kiselev
  et~al\mbox{.}}{2016}]%
        {Kiselev2016-pc}
\bibfield{author}{\bibinfo{person}{Ilya Kiselev}, \bibinfo{person}{Daniel
  Neil}, {and} \bibinfo{person}{Shih~Chii Liu}.}
  \bibinfo{year}{2016}\natexlab{}.
\newblock \showarticletitle{{Event-Driven} Deep Neural Network Hardware System
  for Sensor Fusion}. In \bibinfo{booktitle}{\emph{Proceedings - {IEEE}
  International Symposium on Circuits and Systems}}. \bibinfo{publisher}{IEEE},
  \bibinfo{address}{Montreal, Canada}, \bibinfo{pages}{2495--2498}.
\newblock


\bibitem[\protect\citeauthoryear{LeCun, Boser, Denker, Henderson, Howard,
  Hubbard, and Jackel}{LeCun et~al\mbox{.}}{1989}]%
        {LeCun1989-mu}
\bibfield{author}{\bibinfo{person}{Y LeCun}, \bibinfo{person}{B Boser},
  \bibinfo{person}{J~S Denker}, \bibinfo{person}{D Henderson},
  \bibinfo{person}{R~E Howard}, \bibinfo{person}{W Hubbard}, {and}
  \bibinfo{person}{L~D Jackel}.} \bibinfo{year}{1989}\natexlab{}.
\newblock \showarticletitle{Backpropagation Applied to Handwritten Zip Code
  Recognition}.
\newblock \bibinfo{journal}{\emph{Neural Comput.}} \bibinfo{volume}{1},
  \bibinfo{number}{4} (\bibinfo{year}{1989}), \bibinfo{pages}{541--551}.
\newblock


\bibitem[\protect\citeauthoryear{Lin, Wild, Chinya, Cao, Davies, Lavery, and
  Wang}{Lin et~al\mbox{.}}{2018}]%
        {Lin2018-fg}
\bibfield{author}{\bibinfo{person}{Chit~Kwan Lin}, \bibinfo{person}{Andreas
  Wild}, \bibinfo{person}{Gautham~N Chinya}, \bibinfo{person}{Yongqiang Cao},
  \bibinfo{person}{Mike Davies}, \bibinfo{person}{Daniel~M Lavery}, {and}
  \bibinfo{person}{Hong Wang}.} \bibinfo{year}{2018}\natexlab{}.
\newblock \showarticletitle{Programming Spiking Neural Networks on Intel's
  Loihi}.
\newblock \bibinfo{journal}{\emph{Computer}} \bibinfo{volume}{51},
  \bibinfo{number}{3} (\bibinfo{year}{2018}), \bibinfo{pages}{52--61}.
\newblock


\bibitem[\protect\citeauthoryear{Massa, Marchisio, Martina, and Shafique}{Massa
  et~al\mbox{.}}{2020}]%
        {Massa2020-wd}
\bibfield{author}{\bibinfo{person}{Riccardo Massa}, \bibinfo{person}{Alberto
  Marchisio}, \bibinfo{person}{Maurizio Martina}, {and}
  \bibinfo{person}{Muhammad Shafique}.} \bibinfo{year}{2020}\natexlab{}.
\newblock \showarticletitle{An Efficient Spiking Neural Network for Recognizing
  Gestures with a {DVS} Camera on the Loihi Neuromorphic Processor}. In
  \bibinfo{booktitle}{\emph{2020 International Joint Conference on Neural
  Networks ({IJCNN})}}. \bibinfo{publisher}{IEEE}, \bibinfo{address}{Glasgow,
  Scotland}, \bibinfo{pages}{1}.
\newblock


\bibitem[\protect\citeauthoryear{Merolla, Arthur, Alvarez-Icaza, Cassidy,
  Sawada, Akopyan, Jackson, Imam, Guo, Nakamura, Brezzo, Vo, Esser, Appuswamy,
  Taba, Amir, Flickner, Risk, Manohar, and Modha}{Merolla
  et~al\mbox{.}}{2014}]%
        {Merolla2014-dt}
\bibfield{author}{\bibinfo{person}{P~A Merolla}, \bibinfo{person}{J~V Arthur},
  \bibinfo{person}{R Alvarez-Icaza}, \bibinfo{person}{A~S Cassidy},
  \bibinfo{person}{J Sawada}, \bibinfo{person}{F Akopyan}, \bibinfo{person}{B~L
  Jackson}, \bibinfo{person}{N Imam}, \bibinfo{person}{C Guo},
  \bibinfo{person}{Y Nakamura}, \bibinfo{person}{B Brezzo}, \bibinfo{person}{I
  Vo}, \bibinfo{person}{S~K Esser}, \bibinfo{person}{R Appuswamy},
  \bibinfo{person}{B Taba}, \bibinfo{person}{A Amir}, \bibinfo{person}{M~D
  Flickner}, \bibinfo{person}{W~P Risk}, \bibinfo{person}{R Manohar}, {and}
  \bibinfo{person}{D~S Modha}.} \bibinfo{year}{2014}\natexlab{}.
\newblock \showarticletitle{A Million {Spiking-Neuron} Integrated Circuit With
  a Scalable Communication Network and Interface}.
\newblock \bibinfo{journal}{\emph{Science}} \bibinfo{volume}{345},
  \bibinfo{number}{6197} (\bibinfo{year}{2014}), \bibinfo{pages}{668--673}.
\newblock


\bibitem[\protect\citeauthoryear{Mostafa}{Mostafa}{2018}]%
        {Mostafa2018-st}
\bibfield{author}{\bibinfo{person}{Hesham Mostafa}.}
  \bibinfo{year}{2018}\natexlab{}.
\newblock \showarticletitle{Supervised Learning Based on Temporal Coding in
  Spiking Neural Networks}.
\newblock \bibinfo{journal}{\emph{IEEE Transactions on Neural Networks and
  Learning Systems}} \bibinfo{volume}{29}, \bibinfo{number}{7}
  (\bibinfo{year}{2018}), \bibinfo{pages}{3227--3235}.
\newblock


\bibitem[\protect\citeauthoryear{Mostafa, Pedroni, Sheik, and
  Cauwenberghs}{Mostafa et~al\mbox{.}}{2017}]%
        {Mostafa2017-xr}
\bibfield{author}{\bibinfo{person}{Hesham Mostafa}, \bibinfo{person}{Bruno~U
  Pedroni}, \bibinfo{person}{Sadique Sheik}, {and} \bibinfo{person}{Gert
  Cauwenberghs}.} \bibinfo{year}{2017}\natexlab{}.
\newblock \showarticletitle{Fast Classification Using Sparsely Active Spiking
  Networks}. In \bibinfo{booktitle}{\emph{Proceedings - {IEEE} International
  Symposium on Circuits and Systems}}. \bibinfo{publisher}{IEEE},
  \bibinfo{address}{Baltimore, MD, USA}, \bibinfo{pages}{1}.
\newblock


\bibitem[\protect\citeauthoryear{Orchard, Jayawant, Cohen, and Thakor}{Orchard
  et~al\mbox{.}}{2015}]%
        {Orchard2015-fg}
\bibfield{author}{\bibinfo{person}{Garrick Orchard}, \bibinfo{person}{Ajinkya
  Jayawant}, \bibinfo{person}{Gregory~K Cohen}, {and} \bibinfo{person}{Nitish
  Thakor}.} \bibinfo{year}{2015}\natexlab{}.
\newblock \showarticletitle{Converting Static Image Datasets to Spiking
  Neuromorphic Datasets Using Saccades}.
\newblock \bibinfo{journal}{\emph{Front. Neurosci.}}  \bibinfo{volume}{9}
  (\bibinfo{year}{2015}), \bibinfo{pages}{1}.
\newblock


\bibitem[\protect\citeauthoryear{Paszke, Gross, Massa, Lerer, Bradbury, Chanan,
  Killeen, Lin, Gimelshein, Antiga, Desmaison, Kopf, Yang, DeVito, Raison,
  Tejani, Chilamkurthy, Steiner, Fang, Bai, and Chintala}{Paszke
  et~al\mbox{.}}{2019}]%
        {Paszke2019-ax}
\bibfield{author}{\bibinfo{person}{Adam Paszke}, \bibinfo{person}{Sam Gross},
  \bibinfo{person}{Francisco Massa}, \bibinfo{person}{Adam Lerer},
  \bibinfo{person}{James Bradbury}, \bibinfo{person}{Gregory Chanan},
  \bibinfo{person}{Trevor Killeen}, \bibinfo{person}{Zeming Lin},
  \bibinfo{person}{Natalia Gimelshein}, \bibinfo{person}{Luca Antiga},
  \bibinfo{person}{Alban Desmaison}, \bibinfo{person}{Andreas Kopf},
  \bibinfo{person}{Edward Yang}, \bibinfo{person}{Zachary DeVito},
  \bibinfo{person}{Martin Raison}, \bibinfo{person}{Alykhan Tejani},
  \bibinfo{person}{Sasank Chilamkurthy}, \bibinfo{person}{Benoit Steiner},
  \bibinfo{person}{Lu Fang}, \bibinfo{person}{Junjie Bai}, {and}
  \bibinfo{person}{Soumith Chintala}.} \bibinfo{year}{2019}\natexlab{}.
\newblock \showarticletitle{{PyTorch}: An Imperative Style, {High-Performance}
  Deep Learning Library}. In \bibinfo{booktitle}{\emph{Advances in Neural
  Information Processing Systems 32}},
  \bibfield{editor}{\bibinfo{person}{H~Wallach},
  \bibinfo{person}{H~Larochelle}, \bibinfo{person}{A~Beygelzimer},
  \bibinfo{person}{F~d'Alche Buc}, \bibinfo{person}{E~Fox}, {and}
  \bibinfo{person}{R~Garnett}} (Eds.). \bibinfo{publisher}{Curran Associates,
  Inc.}, \bibinfo{address}{Vancouver, Canada}, \bibinfo{pages}{8026--8037}.
\newblock


\bibitem[\protect\citeauthoryear{Pati{\~n}o-Saucedo, Rostro-Gonzalez,
  Serrano-Gotarredona, and Linares-Barranco}{Pati{\~n}o-Saucedo
  et~al\mbox{.}}{2020}]%
        {Patino-Saucedo2020-mx}
\bibfield{author}{\bibinfo{person}{Alberto Pati{\~n}o-Saucedo},
  \bibinfo{person}{Horacio Rostro-Gonzalez}, \bibinfo{person}{Teresa
  Serrano-Gotarredona}, {and} \bibinfo{person}{Bernab{\'e} Linares-Barranco}.}
  \bibinfo{year}{2020}\natexlab{}.
\newblock \showarticletitle{{Event-Driven} Implementation of Deep Spiking
  Convolutional Neural Networks for Supervised Classification Using the
  {SpiNNaker} Neuromorphic Platform}.
\newblock \bibinfo{journal}{\emph{Neural Netw.}}  \bibinfo{volume}{121}
  (\bibinfo{year}{2020}), \bibinfo{pages}{319--328}.
\newblock


\bibitem[\protect\citeauthoryear{Rathi, Srinivasan, Panda, and Roy}{Rathi
  et~al\mbox{.}}{2020}]%
        {Rathi2020-lt}
\bibfield{author}{\bibinfo{person}{Nitin Rathi},
  \bibinfo{person}{Gopalakrishnan Srinivasan}, \bibinfo{person}{Priyadarshini
  Panda}, {and} \bibinfo{person}{Kaushik Roy}.}
  \bibinfo{year}{2020}\natexlab{}.
\newblock \showarticletitle{Enabling Deep Spiking Neural Networks with Hybrid
  Conversion and Spike Timing Dependent Backpropagation}. In
  \bibinfo{booktitle}{\emph{International Conference on Learning
  Representations}}. \bibinfo{publisher}{OpenReview.net},
  \bibinfo{address}{Addis Ababa, Ethiopia}, \bibinfo{pages}{1}.
\newblock


\bibitem[\protect\citeauthoryear{Reuther, Michaleas, Jones, Gadepally, Samsi,
  and Kepner}{Reuther et~al\mbox{.}}{2019}]%
        {Reuther2019-zj}
\bibfield{author}{\bibinfo{person}{Albert Reuther}, \bibinfo{person}{Peter
  Michaleas}, \bibinfo{person}{Michael Jones}, \bibinfo{person}{Vijay
  Gadepally}, \bibinfo{person}{Siddharth Samsi}, {and} \bibinfo{person}{Jeremy
  Kepner}.} \bibinfo{year}{2019}\natexlab{}.
\newblock \showarticletitle{Survey and Benchmarking of Machine Learning
  Accelerators}. In \bibinfo{booktitle}{\emph{2019 {IEEE} High Performance
  Extreme Computing Conference}}. \bibinfo{publisher}{IEEE},
  \bibinfo{address}{Waltham, MA, USA}, \bibinfo{pages}{1--9}.
\newblock


\bibitem[\protect\citeauthoryear{Rhodes, Bogdan, Brenninkmeijer, Davidson,
  Fellows, Gait, Lester, Mikaitis, Plana, Rowley, Stokes, and Furber}{Rhodes
  et~al\mbox{.}}{2018}]%
        {Rhodes2018-wp}
\bibfield{author}{\bibinfo{person}{Oliver Rhodes}, \bibinfo{person}{Petru{\c
  t}~A Bogdan}, \bibinfo{person}{Christian Brenninkmeijer},
  \bibinfo{person}{Simon Davidson}, \bibinfo{person}{Donal Fellows},
  \bibinfo{person}{Andrew Gait}, \bibinfo{person}{David~R Lester},
  \bibinfo{person}{Mantas Mikaitis}, \bibinfo{person}{Luis~A Plana},
  \bibinfo{person}{Andrew G~D Rowley}, \bibinfo{person}{Alan~B Stokes}, {and}
  \bibinfo{person}{Steve~B Furber}.} \bibinfo{year}{2018}\natexlab{}.
\newblock \showarticletitle{Spynnaker: A Software Package for Running Pynn
  Simulations on Spinnaker}.
\newblock \bibinfo{journal}{\emph{Front. Neurosci.}} \bibinfo{volume}{12},
  \bibinfo{number}{November} (\bibinfo{year}{2018}), \bibinfo{pages}{1}.
\newblock


\bibitem[\protect\citeauthoryear{Rowley, Brenninkmeijer, Davidson, Fellows,
  Gait, Lester, Plana, Rhodes, Stokes, and Furber}{Rowley
  et~al\mbox{.}}{2019}]%
        {Rowley2019-qf}
\bibfield{author}{\bibinfo{person}{Andrew G~D Rowley},
  \bibinfo{person}{Christian Brenninkmeijer}, \bibinfo{person}{Simon Davidson},
  \bibinfo{person}{Donal Fellows}, \bibinfo{person}{Andrew Gait},
  \bibinfo{person}{David~R Lester}, \bibinfo{person}{Luis~A Plana},
  \bibinfo{person}{Oliver Rhodes}, \bibinfo{person}{Alan~B Stokes}, {and}
  \bibinfo{person}{Steve~B Furber}.} \bibinfo{year}{2019}\natexlab{}.
\newblock \showarticletitle{{SpiNNTools}: The Execution Engine for the
  {SpiNNaker} Platform}.
\newblock \bibinfo{journal}{\emph{Front. Neurosci.}}  \bibinfo{volume}{13}
  (\bibinfo{date}{March} \bibinfo{year}{2019}), \bibinfo{pages}{1}.
\newblock


\bibitem[\protect\citeauthoryear{Rueckauer and Liu}{Rueckauer and Liu}{2018}]%
        {Rueckauer2018-sz}
\bibfield{author}{\bibinfo{person}{Bodo Rueckauer} {and}
  \bibinfo{person}{Shih~Chii Liu}.} \bibinfo{year}{2018}\natexlab{}.
\newblock \showarticletitle{Conversion of Analog to Spiking Neural Networks
  using Sparse Temporal Coding}. In \bibinfo{booktitle}{\emph{Proceedings -
  {IEEE} International Symposium on Circuits and Systems}}.
  \bibinfo{publisher}{IEEE}, \bibinfo{address}{Florence, Italy},
  \bibinfo{pages}{8--12}.
\newblock


\bibitem[\protect\citeauthoryear{Rueckauer, Lungu, Hu, Pfeiffer, and
  Liu}{Rueckauer et~al\mbox{.}}{2017}]%
        {Rueckauer2017-ch}
\bibfield{author}{\bibinfo{person}{Bodo Rueckauer},
  \bibinfo{person}{Iulia-Alexandra Lungu}, \bibinfo{person}{Yuhuang Hu},
  \bibinfo{person}{Michael Pfeiffer}, {and} \bibinfo{person}{Shih-Chii Liu}.}
  \bibinfo{year}{2017}\natexlab{}.
\newblock \showarticletitle{Conversion of {Continuous-Valued} Deep Networks to
  Efficient {Event-Driven} Networks for Image Classification}.
\newblock \bibinfo{journal}{\emph{Front. Neurosci.}} \bibinfo{volume}{11},
  \bibinfo{number}{December} (\bibinfo{year}{2017}), \bibinfo{pages}{1--12}.
\newblock


\bibitem[\protect\citeauthoryear{Schmitt, Klahn, Bellec, Grubl, Guttler,
  Hartel, Hartmann, Husmann, Husmann, Jeltsch, Karasenko, Kleider, Koke,
  Kononov, Mauch, Muller, Muller, Partzsch, Petrovici, Schiefer, Scholze,
  Thanasoulis, Vogginger, Legenstein, Maass, Mayr, Schuffny, Schemmel, and
  Meier}{Schmitt et~al\mbox{.}}{2017}]%
        {Schmitt2017-cq}
\bibfield{author}{\bibinfo{person}{Sebastian Schmitt}, \bibinfo{person}{Johann
  Klahn}, \bibinfo{person}{Guillaume Bellec}, \bibinfo{person}{Andreas Grubl},
  \bibinfo{person}{Maurice Guttler}, \bibinfo{person}{Andreas Hartel},
  \bibinfo{person}{Stephan Hartmann}, \bibinfo{person}{Dan Husmann},
  \bibinfo{person}{Kai Husmann}, \bibinfo{person}{Sebastian Jeltsch},
  \bibinfo{person}{Vitali Karasenko}, \bibinfo{person}{Mitja Kleider},
  \bibinfo{person}{Christoph Koke}, \bibinfo{person}{Alexander Kononov},
  \bibinfo{person}{Christian Mauch}, \bibinfo{person}{Eric Muller},
  \bibinfo{person}{Paul Muller}, \bibinfo{person}{Johannes Partzsch},
  \bibinfo{person}{Mihai~A Petrovici}, \bibinfo{person}{Stefan Schiefer},
  \bibinfo{person}{Stefan Scholze}, \bibinfo{person}{Vasilis Thanasoulis},
  \bibinfo{person}{Bernhard Vogginger}, \bibinfo{person}{Robert Legenstein},
  \bibinfo{person}{Wolfgang Maass}, \bibinfo{person}{Christian Mayr},
  \bibinfo{person}{Rene Schuffny}, \bibinfo{person}{Johannes Schemmel}, {and}
  \bibinfo{person}{Karlheinz Meier}.} \bibinfo{year}{2017}\natexlab{}.
\newblock \showarticletitle{Neuromorphic Hardware in the Loop: Training a Deep
  Spiking Network on the {BrainScaleS} {Wafer-Scale} System}. In
  \bibinfo{booktitle}{\emph{Proceedings of the International Joint Conference
  on Neural Networks}}. \bibinfo{publisher}{IEEE}, \bibinfo{address}{Anchorage,
  Alaska, USA}, \bibinfo{pages}{2227--2234}.
\newblock


\bibitem[\protect\citeauthoryear{Shrestha and Orchard}{Shrestha and
  Orchard}{2018}]%
        {Shrestha2018-io}
\bibfield{author}{\bibinfo{person}{Sumit~Bam Shrestha} {and}
  \bibinfo{person}{Garrick Orchard}.} \bibinfo{year}{2018}\natexlab{}.
\newblock \showarticletitle{Slayer: Spike Layer Error Reassignment in Time}. In
  \bibinfo{booktitle}{\emph{Advances in Neural Information Processing
  Systems}}. \bibinfo{publisher}{Curran Associates, Inc.},
  \bibinfo{address}{Montreal, Canada}, \bibinfo{pages}{1412--1421}.
\newblock


\bibitem[\protect\citeauthoryear{Sorbaro, Liu, Bortone, and Sheik}{Sorbaro
  et~al\mbox{.}}{2020}]%
        {Sorbaro2020-dq}
\bibfield{author}{\bibinfo{person}{Martino Sorbaro}, \bibinfo{person}{Qian
  Liu}, \bibinfo{person}{Massimo Bortone}, {and} \bibinfo{person}{Sadique
  Sheik}.} \bibinfo{year}{2020}\natexlab{}.
\newblock \showarticletitle{Optimizing the Energy Consumption of Spiking Neural
  Networks for Neuromorphic Applications}.
\newblock \bibinfo{journal}{\emph{Front. Neurosci.}}  \bibinfo{volume}{14}
  (\bibinfo{year}{2020}), \bibinfo{pages}{662}.
\newblock


\bibitem[\protect\citeauthoryear{Stromatias, Neil, Galluppi, Pfeiffer, Liu, and
  Furber}{Stromatias et~al\mbox{.}}{2015}]%
        {Stromatias2015-kl}
\bibfield{author}{\bibinfo{person}{Evangelos Stromatias},
  \bibinfo{person}{Daniel Neil}, \bibinfo{person}{Francesco Galluppi},
  \bibinfo{person}{Michael Pfeiffer}, \bibinfo{person}{Shih~Chii Liu}, {and}
  \bibinfo{person}{Steve Furber}.} \bibinfo{year}{2015}\natexlab{}.
\newblock \showarticletitle{Scalable {Energy-Efficient}, {Low-Latency}
  Implementations of Trained Spiking Deep Belief Networks on {SpiNNaker}}. In
  \bibinfo{booktitle}{\emph{Proceedings of the International Joint Conference
  on Neural Networks}}. \bibinfo{publisher}{IEEE}, \bibinfo{address}{Killarney,
  Ireland}, \bibinfo{pages}{1--8}.
\newblock


\bibitem[\protect\citeauthoryear{Yousefzadeh, Serrano-Gotarredona,
  Linares-Barranco, Khoei, Hosseini, Holanda, Leroux, Moreira, Tapson, Dhoedt,
  and Simoens}{Yousefzadeh et~al\mbox{.}}{2019}]%
        {Yousefzadeh2019-dd}
\bibfield{author}{\bibinfo{person}{Amirreza Yousefzadeh},
  \bibinfo{person}{Teresa Serrano-Gotarredona}, \bibinfo{person}{Bernabe
  Linares-Barranco}, \bibinfo{person}{Mina~A Khoei}, \bibinfo{person}{Sahar
  Hosseini}, \bibinfo{person}{Priscila Holanda}, \bibinfo{person}{Sam Leroux},
  \bibinfo{person}{Orlando Moreira}, \bibinfo{person}{Jonathan Tapson},
  \bibinfo{person}{Bart Dhoedt}, {and} \bibinfo{person}{Pieter Simoens}.}
  \bibinfo{year}{2019}\natexlab{}.
\newblock \showarticletitle{Asynchronous Spiking Neurons, the Natural Key to
  Exploit Temporal Sparsity}.
\newblock \bibinfo{journal}{\emph{IEEE Journal on Emerging and Selected Topics
  in Circuits and Systems}} \bibinfo{volume}{9}, \bibinfo{number}{4}
  (\bibinfo{date}{Dec.} \bibinfo{year}{2019}), \bibinfo{pages}{668--678}.
\newblock


\bibitem[\protect\citeauthoryear{Yu, Ma, Song, Zhang, Dang, and Tan}{Yu
  et~al\mbox{.}}{2020}]%
        {Yu2020-xk}
\bibfield{author}{\bibinfo{person}{Qiang Yu}, \bibinfo{person}{Chenxiang Ma},
  \bibinfo{person}{Shiming Song}, \bibinfo{person}{Gaoyan Zhang},
  \bibinfo{person}{Jianwu Dang}, {and} \bibinfo{person}{Kay~Chen Tan}.}
  \bibinfo{year}{2020}\natexlab{}.
\newblock \bibinfo{title}{Constructing Accurate and Efficient Deep Spiking
  Neural Networks with Double-threshold and Augmented Schemes}.
  (\bibinfo{date}{May} \bibinfo{year}{2020}).
\newblock


\bibitem[\protect\citeauthoryear{Zambrano, Nusselder, Scholte, and
  Bohte}{Zambrano et~al\mbox{.}}{2019}]%
        {Zambrano2019-ru}
\bibfield{author}{\bibinfo{person}{Davide Zambrano}, \bibinfo{person}{Roeland
  Nusselder}, \bibinfo{person}{H~Steven Scholte}, {and} \bibinfo{person}{Sander
  Bohte}.} \bibinfo{year}{2019}\natexlab{}.
\newblock \showarticletitle{Efficient Computation in Adaptive Artificial
  Spiking Neural Networks}.
\newblock \bibinfo{journal}{\emph{Front. Neurosci.}} \bibinfo{volume}{12},
  \bibinfo{number}{January} (\bibinfo{year}{2019}), \bibinfo{pages}{1--11}.
\newblock


\end{thebibliography}

\end{document}